\pdfoutput=1

\documentclass[a4paper,fleqn,usenatbib]{mn2e}
\usepackage{graphicx}
\usepackage{float}
\usepackage{subfigure}
\usepackage{amssymb}
\usepackage{amsmath}
\usepackage{afterpage}
\usepackage[usenames]{color}
\usepackage[totalwidth=485pt, totalheight=680pt]{geometry}
\usepackage{yfonts}
\usepackage{mathrsfs}
\usepackage{upgreek}
\usepackage{lastpage}

\usepackage{hyperref}
\usepackage[all]{hypcap}

\title{The Faber--Jackson relation and Fundamental Plane from halo abundance matching}
\author[H.~Desmond and R.~H.~Wechsler]{Harry Desmond\thanks{E-mail: harryd2@stanford.edu}
and Risa H.~Wechsler\\
Kavli Institute for Particle Astrophysics and Cosmology and Physics Department, Stanford University, Stanford, CA 94305, USA; \\
SLAC National Accelerator Laboratory, Menlo Park, CA 94025, USA
}

\pubyear{2016}

\begin{document}
\label{FirstPage}
\pagerange{\pageref{FirstPage}--\pageref{LastPage}}
\maketitle

\begin{abstract}
The Fundamental Plane (FP) describes the relation between the stellar mass, size, and velocity dispersion of elliptical galaxies; the Faber--Jackson relation (FJR) is its projection onto \{mass, velocity\} space. In this work we redeploy and expand the framework of~\citet{DW} to ask whether abundance matching-based $\Lambda$CDM models that have shown success in matching the spatial distribution of galaxies are also capable of explaining key properties of the FJR and FP, including their scatter. Within our framework, agreement with the normalisation of the FJR requires haloes to expand in response to disc formation. We find that the tilt of the FP may be explained by a combination of the observed non-homology in galaxy structure and the variation in mass-to-light ratio produced by abundance matching with a universal initial mass function (IMF), provided that the anisotropy of stellar motions is taken into account. However, the predicted scatter around the FP is considerably increased by situating galaxies in cosmologically-motivated haloes due to variations in halo properties at fixed stellar mass, and appears to exceed that of the data. This implies that additional correlations between galaxy and halo variables may be required to fully reconcile these models with elliptical galaxy scaling relations.
\end{abstract}


\section{INTRODUCTION}
\label{sec:intro}

The global dynamical properties of galaxies can be summarised by scaling relations between their mass or luminosity, their size, and the velocity of their constituent stars and gas. For spirals, the Tully--Fisher relation (TFR) describes a tight correlation between galaxy mass and rotation velocity, with a scatter of around 0.08 dex in velocity when stellar mass is used as the independent variable~\citep{P07}, and 0.03 dex for baryonic mass~\citep*{Lelli}. Although a further correlation exists between mass and radius, with a scatter of around 0.17 dex~\citep{P07, Kravtsov13}, it appears that no relation tighter than the TFR can be constructed by including radius information. For ellipticals, however, the analogue of the TFR, the Faber--Jackson relation (FJR;~\citealt{FJ}), \emph{does} exhibit a significant correlation of its residuals with galaxy size, in the sense that smaller than average galaxies have larger than average velocity dispersion. In other words, there exists a relation tighter than the FJR in the form of a Fundamental Plane (FP): $\sigma \propto L^a R^b$~\citep{Djorgovski, Dressler, Bender}. The best-fitting relation has $a \approx 0.8$, $b \approx -0.8$, and a scatter of around 0.09 dex in $R$~\citep{Cappellari_2}.

Since the internal motions of galaxies are set largely by their dark matter mass, these relationships provide key insight into the connection between galaxies and their host haloes. In this work, we construct a semi-empirical framework in $\Lambda$CDM to predict the velocity dispersions of early-type galaxies, and hence the FJR and FP. In particular, our modelling approach is based on subhalo abundance matching (AM), which posits a nearly one-to-one relation between stellar mass and a halo proxy, typically a mass or velocity at a chosen epoch. Models based on this technique have been shown to provide an excellent fit to a wide range of spatial statistics of the galaxy population, including the galaxy-galaxy two and three-point correlation functions, galaxy-galaxy lensing, and group statistics~\citep{Kravtsov,Conroy,Marin08,Behroozi_2010,Guo,Moster,Reddick}. However, AM as it is usually implemented does not distinguish between galaxies of different morphology, and therefore requires testing against the scaling relations of both late and early-type galaxies. Here we investigate whether the properties of the FJR and FP are consistent with the galaxy--halo connection implied by AM. We intend in this way to evaluate agreement between galaxy-scale and large-scale constraints on this connection, and bring out the additional information contained in the FJR and FP on the processes governing galaxy formation.

Our approach may be usefully situated at the intersection of two broad classes of methodology for understanding the dynamical relations of elliptical galaxies. A bottom-up approach focuses on detailed kinematic analysis and mass modelling of a small number of individual galaxies (e.g.~\citealt{Tiret_Combes_MOND_VD, Napolitano, Cappellari_2}). The Jeans equation is solved in full, the anisotropy profile may be determined simultaneously with the velocity dispersion profile, and dark matter fractions in the inner regions are estimated by fitting one or more halo profiles to the data. This approach has the advantage of affording tight control over the model baryonic and dark matter masses of individual systems and hence enables a high precision comparison of theory to data. However, the samples analysed in this way tend to be small (up to a few hundred galaxies), and may therefore be unrepresentative in relevant galaxy properties. Furthermore, dark matter modelling on a galaxy-by-galaxy basis prevents comparison with cosmological expectations for the halo population obtained from $N$-body simulations and a complete galaxy formation model. Except in the most extreme cases, therefore, it is difficult to assess statistically the extent to which the results are consistent with $\Lambda$CDM predictions.

The alternative top-down approach uses a fully cosmological framework to derive model values for galaxy observables. This may be attempted in several different ways depending on the theoretical assumptions employed. Hydrodynamical simulations implement physically motivated prescriptions for baryonic physics in order to simulate galaxy formation in situ and concurrently with the evolution of dark matter haloes (e.g.~\citealt{Illustris, MassiveBlack, EAGLE}). Although in principle applying basic physics in the most straightforward way, such simulations are too computationally expensive to generate theoretical galaxy populations of sufficient size to allow robust statistical comparison with observations. This method may therefore be usefully complemented by semi-analytic and empirical models (e.g.~\citealt{Dutton_1, TG, Dutton_13}), which forgo complete treatment of baryonic physics in favour of simple mappings between galaxy and halo properties in order to predict galaxy observables from the outputs of $N$-body simulations. Although potentially dependent on uncertain parametrizations of key properties, these methods in principle allow statistically rigorous comparison of theoretical and observed galaxy populations.

Hydrodynamical, semi-analytical, and empirical methods have the advantage that they permit a direct test of concrete, cosmologically-motivated galaxy formation scenarios using a representative halo population. On the other hand, theoretical results are typically compared with large and often heterogeneous observational data sets without regard to selection effects, the assumptions and approximations used in the galaxy modelling, or the peculiar nature of particular galaxy subsets. The systematic error budget is therefore uncertain and potentially large. Furthermore, it is rare for such studies to simultaneously investigate the dependence of velocity dispersion on galaxy size in the Fundamental Plane, the scaling relations' scatter, or the effect of radial orbit anisotropy.

By combining the precision of the bottom-up approach with the scope and power of the top-down approach, we intend our methodology to permit detailed inferences of cosmological scope. We will use a complete galaxy formation model based on halo abundance matching to create predictions for the FJR and FP, and establish a set of key diagnostics with which to compare the output of the model to a single homogeneous data set with well-defined selection criteria. We prune the observational data set to those galaxies whose structure is compatible with our mass model, use an identical velocity dispersion measure in the theoretical and observed FJR, ensure the distributions and covariances of size, stellar mass and S\'{e}rsic index are the same between theory and data, and explicitly consider the possible impact of radial orbit anisotropy. Furthermore, we pay particular attention to the tilt and scatter of the FP.

We note that an alternative approach which has been used in the literature to model elliptical galaxies is based on the Modified Newtonian Dynamics (MOND) as opposed to $\Lambda$CDM paradigm. These models seek to explain the dynamics of galaxies by positing a breakdown of Newtonian gravity at low acceleration, instead of invoking dark matter haloes.  Work done along these lines includes~\citet{Sanders_FP},~\citet{Sanders_FJR},~\citet{MOND_FP}, and~\citet{MOND_FP_baby}, and suggests that this framework may be able to offer a convincing explanation of the tilt and small scatter of the Fundamental Plane with realistic assumptions about galaxy structure and radial orbit anisotropy. In MOND, an elliptical lies on a Fundamental Plane near the Newtonian virial relation when the acceleration of stars in its inner region is high and hence their dynamics are quasi-Newtonian; the Faber--Jackson relation (of the specific form $M_* \propto \sigma^4$) is the effective virial relation in the low-acceleration deep-MOND regime that obtains in low surface brightness galaxies and the outer regions of high surface brightness ones. However, none of these models have been shown to match the spatial statistics of galaxies on large scales.

The structure of this paper is as follows. Section~\ref{sec:data} describes the observational data against which we test our model, and documents the $N$-body simulation from which halo properties are drawn. Section~\ref{sec:method} details our methodology. In Section~\ref{sec:results} we present our results for the Faber--Jackson relation and Fundamental Plane. We will develop a series of models of increasing complexity, comparing in each case the predicted and observed values of five key properties of the FJR and FP. The main results are summarised in Table~\ref{tab:table} and Figure~\ref{fig:FJR_cont}. In Section~\ref{sec:discussion} we discuss the general significance and broader ramifications of our results, as well as their relation to results in the literature. Section~\ref{sec:conclusion} concludes.

This work may be seen as a companion to~\citet{DW} (hereafter DW15), to which it is similar in approach and methodology. In DW15 we explored the aptitude of our AM framework to reproduce the spiral galaxy Tully--Fisher and mass--size relations; here we redeploy and augment these models in order to investigate the elliptical galaxies that we previously excluded. In several places we refer to DW15 for more detail.

\section{Observed and Simulated Data}
\label{sec:data}

\subsection{Observational data}
\label{sec:obs}

We compare our model to data from the NASA-Sloan Atlas (NSA).\footnote{\url{www.nsatlas.org/}} This is a compilation of galaxy properties within the footprint of SDSS DR8 out to redshift $z \sim 0.05$, cross-matched with a number of other catalogues. Sky-subtraction and photometric determinations in the NSA are believed to be among the most reliable available~\citep{Blanton_NSA}. We will cast all scaling relations in terms of stellar mass rather than luminosity,\footnote{In a slight abuse of terminology, we shall refer to the relation $M_* \propto \sigma^a R^b$ as the Fundamental Plane. This is sometimes called the ``mass plane'' in the literature to distinguish it from the relations $L \propto \sigma^a R^b$ and $\mu \propto \sigma^a R^b$.} and take mass-to-light ratios from the {\tt kcorrect} algorithm~\citep{Blanton_Roweis} with a universal Chabrier IMF (but see Section~\ref{sec:contraction} for a discussion of the effect of IMF choice).

We select from the NSA a subsample of galaxies to compare to our FJR and FP model predictions. First, we clean the data by eliminating galaxies with unreliable size or redshift measurements~\citep{Bernardi_sample}, and restrict to $z<0.05$. We next deduce the angular diameter distance $d_A$, effective radius $R_\mathrm{eff}$, and S\'{e}rsic index $n$ of each galaxy. The angular diameter distance is calculated from the redshift assuming a flat cosmology with $\Omega_\mathrm{m} = 0.3$ and $h = 0.7$. $R_\mathrm{eff}$ is derived by first multiplying the \texttt{SERSIC\_TH50} field of the NSA (the angular half-light radius along the major axis) by the angular diameter distance, and then circularising the light profile by multiplying by the square root of the minor-to-major axis ratio. The spherically-symmetric S\'{e}rsic index $n$ is set equal to the 2D S\'{e}rsic index recorded in the NSA, from which we find it is not systematically offset. We thus determine the relative frequency distribution \{$M_*$, $R_\mathrm{eff}$, $n$, $d_A$\}, which we will use in Section~\ref{sec:model} to assign $R_\mathrm{eff}$, $n$ and $d_A$ values to our model galaxies as a function of $M_*$.\footnote{While stellar masses in the NSA are reported in units of $h^{-2} \: M_\odot$, we use units of $M_\odot$ throughout.}

Next, we impose the following selection criteria to ensure that the galaxies in our final sample have mass distributions that adhere as closely as possible to the spherically-symmetric S\'{e}rsic profile that we will give our model galaxies. These are typical cuts for eliminating rotation-dominated galaxies (e.g.~\citealt{Bernardi_sample, MOND_FP, Dutton_1}).

\begin{enumerate}

\item{} S\'{e}rsic index $2.5 < n < 5.5$ (low-$n$ galaxies are more likely to be flattened and discy, and the SDSS analysis code has a hard upper limit at $n=6$).

\item{} Concentration ratio $R_{90}/R_{50} > 2.5$ (pressure-supported systems tend to be less compact).

\item{} Apparent axis ratio $b / a > 0.6$ (low projected ellipticity favours more spherical early-type galaxies).

\end{enumerate}

We will refer to the galaxies passing these cuts as ``early-type'' or ``elliptical'' and the remainder as ``late-type'' or ``spiral.'' We caution, however, that these definitions may not conform to others in the literature, which may include a cut on colour (e.g.~\citealt{Dutton_1, Dutton_13}), or be based on visual inspection (e.g.~\citealt{Lintott}). Since we will find that the precision of our analysis is not limited by statistical uncertainties due to sample size, we choose the threshold values for our cuts to be fairly stringent to maximise agreement with our model assumptions. Taken together, these factors result in ``elliptical fractions'' in our analysis significantly lower than in many literature studies (e.g.~\citealt{Bamford, Henriques}).

We next impose a stellar mass cut to generate a volume-limited sample with the same $M_*$ distribution as our AM mocks. Assuming a conservative apparent magnitude limit of $m_r = 17.6$, the SDSS and hence NSA data are complete to $M_r-5\log(h) = -18.4$ out to $z=0.05$~\citep{Zehavi}. We find that the stellar mass value above which all galaxies in both the early and late-type subsample\footnote{According to most definitions in the literature, the limiting stellar mass for a given magnitude threshold is different for early and late-type galaxies. We find that the morphological properties that we select on correlate only weakly with colour in the NSA sample, and hence our early and late-type subsets have comparable mass-to-light ratios.} have $M_r-5\log(h) < -18.4$ is $M_* = 10^{10} M_\odot$, which we impose as the lower $M_*$ limit in our analysis.

These cuts leave 6,088 elliptical galaxies out of 36,520 total. Since we will be interested in the possible impact of the elliptical selection criteria on our theoretical galaxy population (Section~\ref{sec:model}), we record also the fraction of galaxies retained as a function of stellar mass. This rises from $\sim 0.05$ at $M_*=10^{10} \mathrm{M_\odot}$ to $\sim 0.35$ at $M_*=10^{11.7} \mathrm{M_\odot}$, the highest stellar mass we consider.

Finally, we generate an early-type catalogue of \{$M_*$, $R_\mathrm{eff}$, $\sigma_\mathrm{ap}$\} values, to which we will compare our model predictions. We include at this stage a final requirement, $70\;\mathrm{km/s} < \sigma_\mathrm{ap} < \mathrm{420\;km/s}$, which removes galaxies whose velocities are untrustworthy~\citep{Bernardi_sample}. We will apply this cut in the same way to the mock data sets. The final early-type subsample contains 5,682 galaxies.

\subsection{Simulated data}
\label{sec:sim}

Our theoretical galaxy population is based on the haloes generated in the \textsc{darksky-400} simulation~\citep{DarkSky}, a $(400 \: \mathrm{Mpc~h^{-1}})^3$ box with $4096^3$ particles, run with the \textsc{2hot} code~\citep{Warren13}. The \textsc{darksky} suite\footnote{\url{http://darksky.slac.stanford.edu/}} assumes a flat $\Lambda$CDM cosmology with $h=0.688$, $\Omega_\mathrm{m} = 0.295$, $n_\mathrm{s} = 0.968$, and $\sigma_8 = 0.834$. The \textsc{darksky-400} box is large enough to contain sufficient haloes at the high end of our stellar mass range to permit robust statistical testing, and at the same time has a small enough particle mass that haloes at the low mass end are well resolved (according, for example, to the criteria of~\citealt{Diemer_Kravtsov}). We identify haloes using the \textsc{rockstar} halo finder~\citep{Rockstar}. The simulation contains around 1,000,000 haloes hosting galaxies in the stellar mass range $10^{10}-10^{11.7} \mathrm{M_\odot}$.

\section{Method}
\label{sec:method}

The overall aim of our method is to create a map between a set of theoretical parameters describing galaxy formation processes and five key properties of the corresponding theoretical FJR and FP. We do this by generating for particular input parameter values a population of galaxies with stellar masses ($M_*$), effective radii ($R_\mathrm{eff}$), and velocity dispersions averaged over the 1.5'' SDSS aperture ($\sigma_\mathrm{ap}$). We then draw mock data sets from these populations and determine their FJR and FP. By comparing to the observed relations, we assess the aptitude of the models to account for elliptical galaxies' velocity dispersions. The scope of our parametrizations encompasses many of the galaxy formation models under consideration today.

\subsection{Model}
\label{sec:model}

We begin by assigning a stellar mass to each halo in the \textsc{darksky-400} simulation by abundance matching to the $z=0$ snapshot~\citep{Kravtsov,Conroy,Behroozi_2010,Guo,Moster,Reddick}. Specifically, we adopt the AM model of~\citet{Lehmann}, which uses a continuous parameter $\alpha$ to interpolate proxy between virial mass ($\alpha=0$) and maximum circular velocity $v_\mathrm{max}$ ($\alpha=1$) at the time of peak mass. A universal Gaussian scatter (which we refer to as the ``AM scatter'') describes the deviation of the stellar mass--proxy relation from exact monotonicity.~\citet{Lehmann} show that a good fit to the correlation function and satellite fraction of SDSS can be obtained in the case $\alpha \approx 0.6$, AM scatter $\approx 0.16$, and we take these as our fiducial values. We use the stellar mass function of~\citet{Bernardi_SMF} based on single-S\'{e}rsic fits, which is consistent with the stellar mass determinations of the NSA.

Each model galaxy is assumed to have a~\citet{Sersic_1968} mass density profile\footnote{We will assume that stellar mass traces $r$-band light, so that the luminosity and mass profiles are directly related. Strong trends in stellar mass-to-light ratio are not observed in elliptical galaxies, and there are only small differences in NSA measurements between bands.}

\begin{equation}
\Sigma(r) = \Sigma_e \exp\left\{-b_n\left[\left(\frac{R}{R_\mathrm{eff}}\right)^{1/n}-1\right]\right\},
\end{equation}

\noindent where $b_n$ is chosen so that the luminosity within $R_\mathrm{eff}$ is half the total luminosity~\citep{Ciotti}

\begin{equation}
b_n \approx 2n - \frac{1}{3} + \frac{0.009876}{n}.
\end{equation}

\noindent A good approximation to the de-projected volume density is~\citep{Prugniel_Simien}

\begin{equation}
\rho(r) = \rho_0 \left(\frac{r}{R_\mathrm{eff}}\right)^{-p_n} \exp\left[-b_n \left(\frac{r}{R_\mathrm{eff}}\right)^{1/n}\right],
\end{equation}

\noindent with

\begin{equation}
\rho_0 = \frac{\Sigma_0 \: b_n^{n(1-p_n)}}{2 R_\mathrm{eff}} \frac{\Gamma(2n)}{\Gamma[n(3-p_n)]}
\end{equation}

\noindent and

\begin{equation}
p_n = 1 - \frac{0.6097}{n} + \frac{0.00563}{n^2}.
\end{equation}

We next assign an effective radius $R_\mathrm{eff}$, S\'{e}rsic index $n$ and angular diameter distance $d_A(z)$ to each model galaxy by randomly drawing from the relative frequency distribution calculated in Section~\ref{sec:obs} at the stellar mass of the galaxy in question. This purely empirical method guarantees that the model galaxies have precisely the same \{$M_*$, $R_\mathrm{eff}$, $n$, $d_A$\} distribution as the data to which we will compare them.\footnote{In DW15 we set size using a theoretical spin-based model along the lines of~\citet{MMW}. There are two reasons to change method here: 1) we showed in DW15 that the spin-based model generates too high a scatter in the mass--size relation, and this may have undesired ramifications for the predicted FJR and FP, and 2) elliptical galaxies have very little rotation and it is not clear to what extent the little angular momentum they do possess is correlated with halo spin -- mass loss and merging may be more important. We discuss the broader consequences of these two size models in Section~\ref{sec:discrepancies}.} For reference, the best-fitting power-law mass--size relation (MSR) is $\log(R_\mathrm{eff}/\mathrm{kpc}) = 0.489 + 0.533(\log(M_*/\mathrm{M_\odot}) - \langle \log(M_*/\mathrm{M_\odot}) \rangle)$, where $\langle \log(M_*/\mathrm{M_\odot}) \rangle = 10.6$ is the median mass of the NSA sample. The scatter in size is 0.136 dex. The angular diameter distance is then used to calculate the physical size of the SDSS aperture at the location of each galaxy, within which the weighted velocity dispersion $\sigma_\mathrm{ap}$ is defined. Combined with present day halo virial mass ($M_\mathrm{vir}$) and concentration, \{$M_*$, $R_\mathrm{eff}$, $n$, $d_A$\} form the inputs to the calculation of $\sigma_\mathrm{ap}$, which we now perform.

For a given cumulative spherical mass distribution $M(r)$, density distribution $\rho(r)$, and radial orbit anisotropy profile $\beta(r) \equiv 1-\sigma_\theta^2(r)/\sigma_r^2(r)$, the radial velocity dispersion $\sigma_r$ obeys the spherical Jeans equation

\begin{equation}
\frac{d[\rho(r) \sigma_r^2(r)]}{dr} + \frac{2 \beta(r)}{r} \rho(r) \sigma_r^2(r) = -\frac{G \rho(r) M(r)}{r^2}.
\end{equation}

\noindent The line-of-sight velocity dispersion

\begin{equation}
\sigma_\mathrm{los}(R) \equiv \sqrt{\frac{2}{\Sigma(R)} \left[\int^\infty_R \frac{\rho(r) \sigma_r^2 r dr}{\sqrt{r^2 - R^2}} - R^2 \int^\infty_R \frac{\beta \rho(r) \sigma_r^2 dr}{r \sqrt{r^2 - R^2}} \right]},
\end{equation}

\noindent is then given by~\citep{Mamon_Lokas}

\begin{equation}
\sigma_\mathrm{los}(R) = \sqrt{\frac{2 G}{\Sigma(R)} \int^\infty_R K(r/R) \rho(r) M(r) \frac{dr}{r}},
\end{equation}

\noindent where

\begin{align} \nonumber
K(u) &\equiv \frac{1}{2} u^{2\beta - 1} \left[ \left(\frac{3}{2} - \beta\right) \sqrt{\pi} \: \frac{\Gamma(\beta - \frac{1}{2})}{\Gamma(\beta)} \right.\\
&+ \left. \beta B\left(\frac{1}{u^2}, \: \beta + \frac{1}{2}, \: \frac{1}{2}\right) - B\left(\frac{1}{u^2}, \: \beta - \frac{1}{2}, \: \frac{1}{2}\right) \right]
\end{align}

\noindent and $B$ is the incomplete beta function. Finally, we calculate the physical radius $R_\mathrm{ap}$ of the SDSS aperture at the position of the galaxy by multiplying the angular radius of 1.5'' by $d_A$. $\sigma_\mathrm{los}$ is then averaged (weighted by brightness) over the area of the aperture to yield the aperture velocity dispersion

\begin{equation}
\sigma_\mathrm{ap} \equiv \sigma(R_\mathrm{ap}) = \sqrt{\frac{\int^{R_\mathrm{ap}}_0 \Sigma(r) \: \sigma_\mathrm{los}^2(r) \: r dr}{\int^{R_\mathrm{ap}}_0 \Sigma(r) \: r dr}}.
\end{equation}

It now remains to specify 1) the dark matter mass distribution $\rho_{DM}(r)$, 2) the radial orbit anisotropy $\beta(r)$, and 3) which haloes in the simulation should be assumed to harbour the elliptical galaxies in question, and hence included in the analysis. Knowing the virial mass and concentration of each halo, we can reconstruct the NFW density profile; however, it is likely that this profile is altered in some way by the formation of galaxies at the halo centres. One possibility is for the dark matter to contract adiabatically~\citep*{Blumenthal, Gnedin_2004, Gnedin_2011}, but it is also possible for baryons to transfer energy to the halo and cause it to expand instead (e.g.~\citealt{HydroSim_cores, diCintio, Pontzen}). We employ here the same parametrization of these effects as in DW15: a free, continuous parameter $\nu$ interpolates between the adiabatic contraction prescription of~\citet{Gnedin_2011} ($\nu=1$), and an expansion of the same magnitude ($\nu = -1$).\footnote{See also~\citealt{D07}, Section 3.2.} $\nu=0$ corresponds to no effect of disc formation, in which case haloes retain their NFW form.

The anisotropy parameter $\beta$ characterises the extent to which stars follow predominantly tangential or radial orbits. $\beta=-\infty$ describes motion that is purely tangential, $\beta=0$ isotropic and $\beta=1$ purely radial. In principle, $\beta$ is not only a function of position within each galaxy, but also varies between galaxies. Canonical studies of anisotropy include~\citet{Gerhard_anisotropy},~\citet{Cappellari_anisotropy} and~\citet{Thomas_anisotropy}, and indicate that $\beta$ typically rises in the central regions of ellipticals, and then flattens out or falls slowly towards larger $r$. Variations among galaxies appear to be relatively small and not significantly correlated with other galaxy properties, and in relatively few cases is $\beta$ observed to fall outside the range [$-0.2$, $0.5$]. We will therefore assume $\beta$ to be independent of $r$ in our models (equivalently, we use a single effective value averaged over the aperture). In the most general case we allow $\beta$ to be a function of stellar mass but uncorrelated with all other galaxy and halo variables, and give it a universal Gaussian scatter $\sigma_\beta$ among galaxies: $\beta \sim \mathcal{N}(f(M_*), \sigma_\beta)$.

In Section~\ref{sec:obs} we calculated the fraction of galaxies in the NSA that are ellipticals according to our selection criteria, as a function of stellar mass. This must equal the fraction of haloes hosting elliptical galaxies, and only these haloes ought to be included in our mock population.\footnote{As discussed in Section~\ref{sec:obs}, the elliptical fraction is a function of our early-type cuts. If less stringent cuts were used, leading to an increase in the elliptical fraction, selection effects would have a weaker impact on the theoretical population. Our model therefore provides an upper limit for the possible effect of selection on the FJR and FP.} Given that there is still significant theoretical uncertainty in which haloes should be included, we adopt the flexible selection model of DW15 (in which the motivation and method is spelled out in detail). In short, we assume early-type galaxies to inhabit haloes that are above average in mass or concentration at a given stellar mass, so that their characteristic velocities are higher. The strength of this correlation is governed by a free parameter we call the ``selection factor'': when this parameter takes its largest allowed value ($\pi/2$) ellipticals are put in the haloes which give the \emph{largest} values for $\sigma_\mathrm{ap}$ at given $M_*$, and when it is minimised ($\arctan(0.5)=0.4636$) the correlation between galaxy morphology and halo type is switched off. We expect from the results of DW15 (in addition to recent indications in data; see e.g.~\citealt{Puebla},~\citealt{Wojtak} and~\citealt{Mandelbaum}) that the selection factor ought to take a value in the upper half of its allowed range. We begin with the value 1.4, corresponding to a moderate increase in predicted velocity for ellipticals over the case of random selection.

This completes the description of the inputs to the model: \{$\alpha$, AM scatter, $\nu$, selection factor\} along with a function $f(M_*)$ for setting $\langle \beta \rangle$. Given values for each of these, we are now in a position to calculate $\sigma_\mathrm{ap}$ for the galaxy in each halo of the elliptical subsample. Since there are $\sim 1,000,000$ haloes, and 16.7 per cent of the NSA galaxies in this mass range were determined to be ellipticals (Section~\ref{sec:obs}), our theoretical parent population contains around 167,000 objects. We now describe our procedure for comparing this population with the observational sample in order to evaluate goodness-of-fit and investigate the significance of the model parameters.

\subsection{Comparison of theory and observation}
\label{sec:comparison}

We begin by using the model to generate a mock data set with as many galaxies as the observational sample and at the same stellar masses, but with radii and velocity dispersions drawn randomly from the theoretical population. For consistency with our observational sample, we impose at this stage the requirement $70\;\mathrm{km/s} < \sigma_\mathrm{ap} < \mathrm{420\;km/s}$.

The stellar masses used in AM as well as the sizes and S\'{e}rsic indices we give our mock galaxies all derive from the observed distributions, and hence no correction need be applied to convert between true and observed values of these quantities when comparing the mock and real data. Measurement uncertainties do however play a role in determining $\sigma^\mathrm{obs}_\mathrm{ap}$ for a given mock galaxy, which is a function of the true rather than observed values of $M_*$, $R_\mathrm{eff}$ and $n$, and furthermore is different to the $\sigma^\mathrm{true}_\mathrm{ap}$ value naturally output by the model. In a rigorously Bayesian analysis, the true positions of the galaxies in \{$M_*$, $R_\mathrm{eff}$, $n$, $\sigma_\mathrm{ap}$\} space would be treated as free parameters with posterior distributions deduced simultaneously with the others. At the maximum likelihood point, the true positions would lie somewhat closer to the model prediction than the observed positions (within the measurement uncertainties), and hence we would find slightly greater agreement between theory and data. However, since this analysis is too expensive to perform here we take the true and observed positions to be coincident. This is justified by the relatively small uncertainties in the measurements of the local NSA galaxies, and is no worse an estimate for the maximum-likelihood values of the true parameters than those derived by randomly scattering the observed values.

We fit to each mock data set a power-law FJR\footnote{Note that the neglect of measurement uncertainties means that the values of our power-law parameters are not commensurable with those that would be derived with measurement errors deconvolved, and that the scatters we quote are not intrinsic. In addition, these parameters are dependent on the particular selection criteria of our sample, especially the $\sigma_\mathrm{ap}>70\;\mathrm{km/s}$ cut. We are not concerned here with the determination of the fitting parameters themselves, but only their comparison between real and mock data sets. This requires merely that each data set be treated in the same way.} of the form

\begin{equation}
\log(\sigma_\mathrm{ap}/\mathrm{km}\;\mathrm{s}^{-1}) = c_{FJR} + m \: \left(\log(M_*/\mathrm{M_\odot}) - \langle \log(M_*/\mathrm{M_\odot}) \rangle \right),
\end{equation}

\noindent with scatter $s_{FJR}$ in $\log(\sigma_\mathrm{ap}/\mathrm{km} \; \mathrm{s}^{-1})$, and a planar FP

\begin{align} \nonumber
\log(M_*/\mathrm{M_\odot}) &= c_{FP} \\ \nonumber
&+ m_V \: \left(\log(\sigma_\mathrm{ap}/\mathrm{km} \; \mathrm{s}^{-1}) - \langle \log(\sigma_\mathrm{ap}/\mathrm{km}\;\mathrm{s}^{-1}) \rangle\right) \\
&+ m_R \: \left(\log(R_\mathrm{eff}/\mathrm{kpc}) - \langle \log(R_\mathrm{eff}/\mathrm{kpc}) \rangle \right)
\end{align}

\noindent with scatter $s_{FP}$ in $\log(M_*)$. We then reduce the set \{$c_{FJR}$, $m$, $s_{FJR}$, $c_{FP}$, $m_V$, $m_R$, $s_{FP}$\} to 5 \emph{independent} parameters by removing duplicate information: the normalisation $c_{FP}$ of the FP contains no information absent from the combination of the normalisation of the MSR (matched by construction) and that of the FJR ($c_{FJR}$), and one linear combination of $m_V$ and $m_R$ replicates the information in the MSR and FJR slopes. The other degree of freedom, $-\frac{m_R}{m_V} \equiv \frac{d\sigma_\mathrm{ap}}{dR_\mathrm{eff}}|_{M_*}$, which we denote by $\kappa$, describes the correlation of the MSR and FJR residuals and is a measure of the FP tilt.  $\boldsymbol{p} \equiv (c_{FJR}, \: m, \: s_{FJR}, \: s_{FP}, \: \kappa)$ will be the diagnostic statistic that we use to test the theory.

By fitting a plane to the \{$M_*$, $R_\mathrm{eff}$, $\sigma_\mathrm{ap}$\} data we neglect the possibility of curvature in the FP and a dependence of $s_{FP}$ or $\kappa$ on $M_*$. Although we have verified by eye that the predicted and observed FPs are not far from planar, we cannot preclude the possibility of a larger measured $s_{FP}$ being due in part to greater curvature. The FP tilt is known to vary with stellar mass~\citep{Dutton_13}, and we have qualitatively identified such a trend in our data sets. In principle this dependence contains further information. We will find, however, that the parameters of our framework are well constrained by fitting an averaged tilt and hence we leave an investigation of the significance of this effect within the AM framework to further work.

We evaluate goodness-of-fit as follows. For each set of parameter values we generate 1000 Monte Carlo \{$M_*$, $R_\mathrm{eff}$, $\sigma_\mathrm{ap}$\} realisations, and calculate for each one the best-fitting value of $\boldsymbol{p}$. We then condense this information into a set of median values $\langle \boldsymbol{p} \rangle$ over the realisations (the values one would be most likely to find in the data if the model were true) and the standard deviation $\boldsymbol{\sigma}_p$ (an indication of the expected spread between different observational samples). We shall find the skews of the Monte Carlo $\boldsymbol{p}$ distributions to be negligible, making them fully characterisable by these two statistics. A measure of goodness-of-fit is then $\frac{\langle \boldsymbol{p} \rangle - \boldsymbol{p}_d}{\langle \boldsymbol{\sigma}_p \rangle}$, where $\boldsymbol{p}_d$ is the vector $\boldsymbol{p}$ for the NSA data. The more standard deviations the real data from the model median, the worse the model is performing.

We will investigate a total of seven different sets of values for the input model parameters, arranged in order of increasing complexity. To begin with, and for reference, we examine the scenario (which we label ``no-DM'') in which velocity dispersions are determined entirely by the baryonic mass, and stellar motions are isotropic. In this case, none of the other model parameters apply. We then introduce dark matter haloes using the prescription of Section~\ref{sec:model}, first with a fiducial model in which the parameters are given values which might be considered most likely a priori (although still with $\beta=0$), then with individual perturbations to the parameters, and finally with parameter values chosen to optimise the fit to the FJR and FP. We then lift the $\beta=0$ restriction and consider a model with small radial anisotropies (favoured by existing data), and one in which the effective anisotropy increases with stellar mass.

For each set of parameter values we show a contour plot of the predicted FJR (Fig.~\ref{fig:FJR_cont}), and list the median $\langle \boldsymbol{p} \rangle$ of the five diagnostics (Table~\ref{tab:table}). In certain cases we show also a contour plot of the correlation of the velocity and radius residuals (Fig.~\ref{fig:Rescorr_cont}), and the standard deviations $\boldsymbol{\sigma}_p$ of the diagnostics among the Monte Carlo realisations (Table~\ref{tab:table2}).

For reference, the free parameters of our framework are listed and described in Table~\ref{tab:params1}, and the statistics of the FJR and FP to which we compare our model are given in Table~\ref{tab:params2}.

\begin{table*}
  \begin{center}
    \begin{tabular}{l|r}
      \hline
      $\alpha$					& Interpolates AM proxy between $M_\mathrm{vir}$ ($\alpha=0$) and $v_\mathrm{max}$ ($\alpha=1$)\\
      AM scatter				& Universal Gaussian scatter in stellar mass at fixed proxy\\
      $\nu$					& Controls degree of halo expansion ($-$) or contraction (+)\\
      selection factor				& Governs impact of selection effects on theoretical FJR and FP\\
      $\mathcal{N}(f(M_*), \sigma_\beta)$	& Distribution from which galaxy anisotropy $\beta$ is drawn\\
      \hline
    \end{tabular}
  \caption{Free parameters of the framework.}
  \label{tab:params1}
  \end{center}
\end{table*}

\begin{table}
  \begin{center}
    \begin{tabular}{l|r}
      \hline
      $m$			& FJR slope\\
      $c_{FJR}$			& FJR normalisation\\
      $s_{FJR}$			& FJR scatter\\
      $\kappa$			& Average $\frac{d\sigma_\mathrm{ap}}{dR_\mathrm{eff}}|_{M_*}$\\
      $s_{FP}$			& FP scatter\\
      \hline
    \end{tabular}
  \caption{Diagnostic statistics (components of $\boldsymbol{p}$) of the FJR and FP. See also Eqs. 11 and 12.}
  \label{tab:params2}
  \end{center}
\end{table}

\section{Results}
\label{sec:results}

The first row of Table~\ref{tab:table} lists the values of the five FJR and FP diagnostic parameters as measured in the data. A few observations are in order before we present our model predictions. First, if elliptical galaxies were homologous and all in perfect virial equilibrium in the absence of dark matter, $\kappa$ would be $-0.5$. The measurement from the data therefore shows that in reality there is a \emph{stronger} dependence of velocity on radius than in this oversimplified case (at least under this IMF choice; see Section~\ref{sec:contraction}). Second, $s_{FJR}$ is a scatter in velocity while $s_{FP}$ is a scatter in mass. The effective scatter in mass from the FJR, 0.304, is obtained by dividing the velocity scatter by the FJR slope. We see that the precision with which one can determine $\log(M_*/\mathrm{M_\odot})$ is approximately doubled by including radius information, i.e. in progressing from the FJR to the FP.

\subsection{No-DM}
\label{sec:noDM}
\begin{figure*}
  \subfigure[No-DM]
  {
    \includegraphics[width=0.395\textwidth]{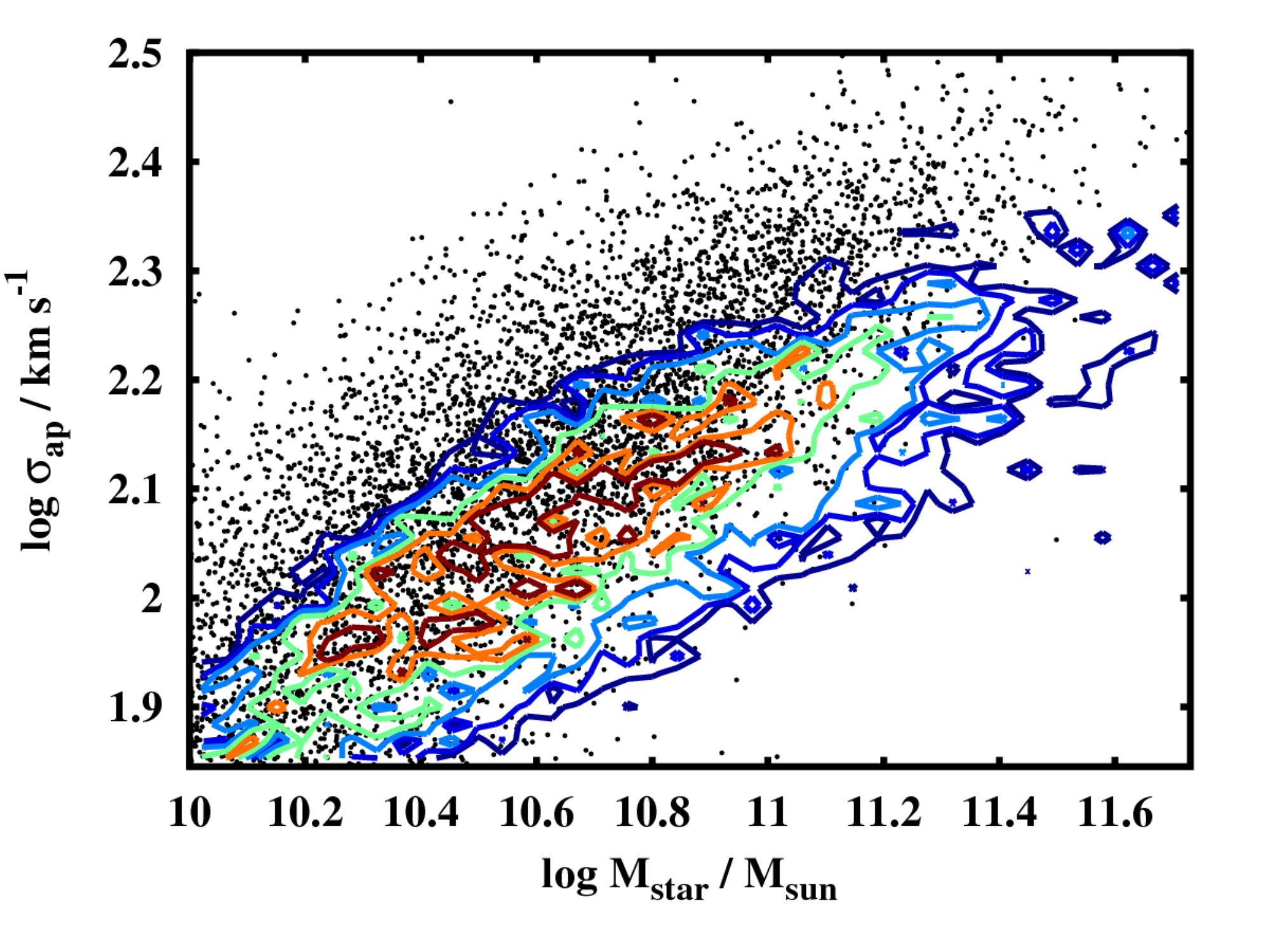}
    \label{fig:FJR_noDM}
  }
  \subfigure[Fiducial: $\alpha=0.6$, AM scatter=0.16, $\nu=0.5$, selection factor=1.4]
  {
    \includegraphics[width=0.395\textwidth]{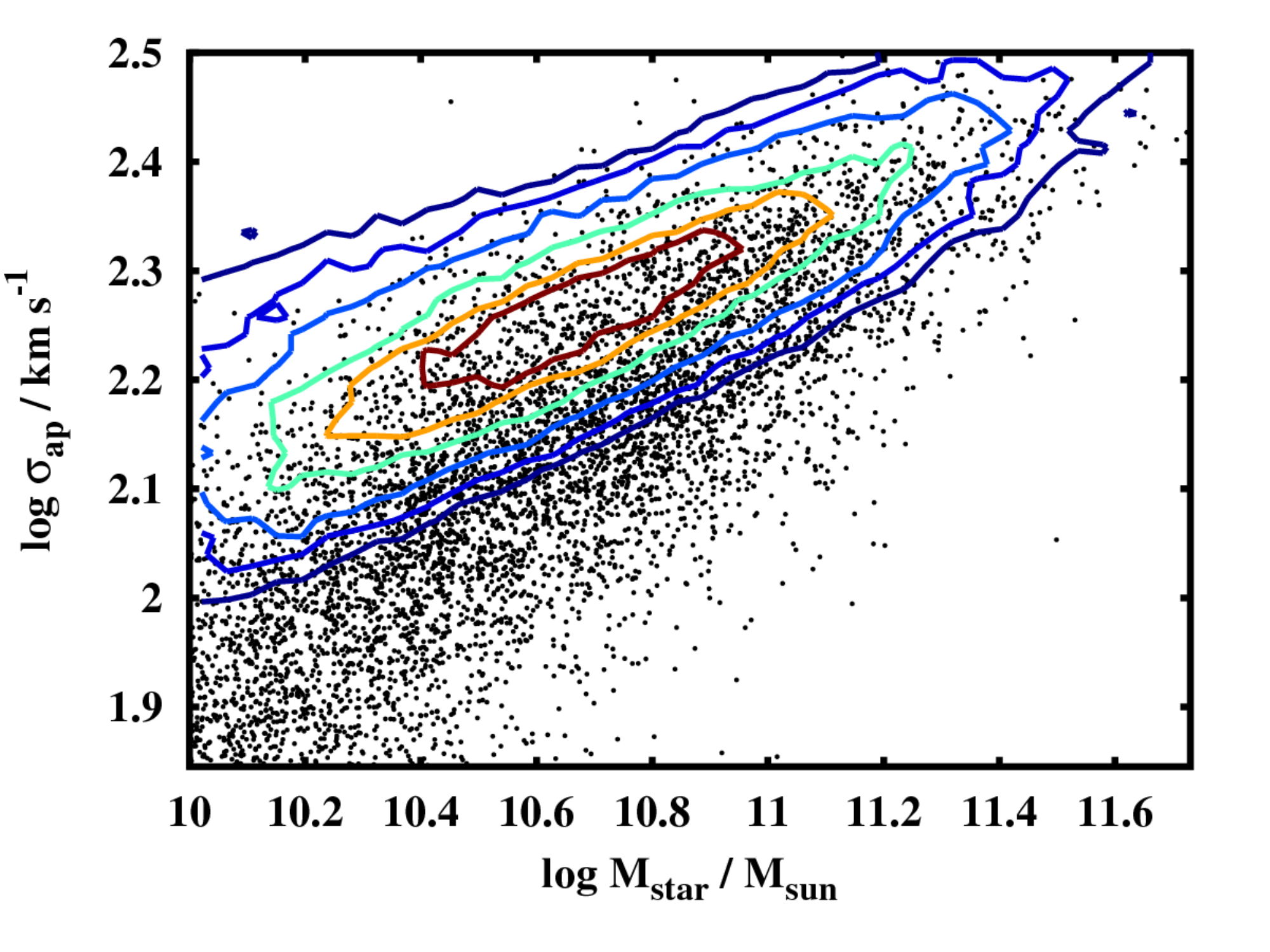}
    \label{fig:FJR_fid}
  }
  \subfigure[AM scatter $\rightarrow$ 0.4]
  {
    \includegraphics[width=0.395\textwidth]{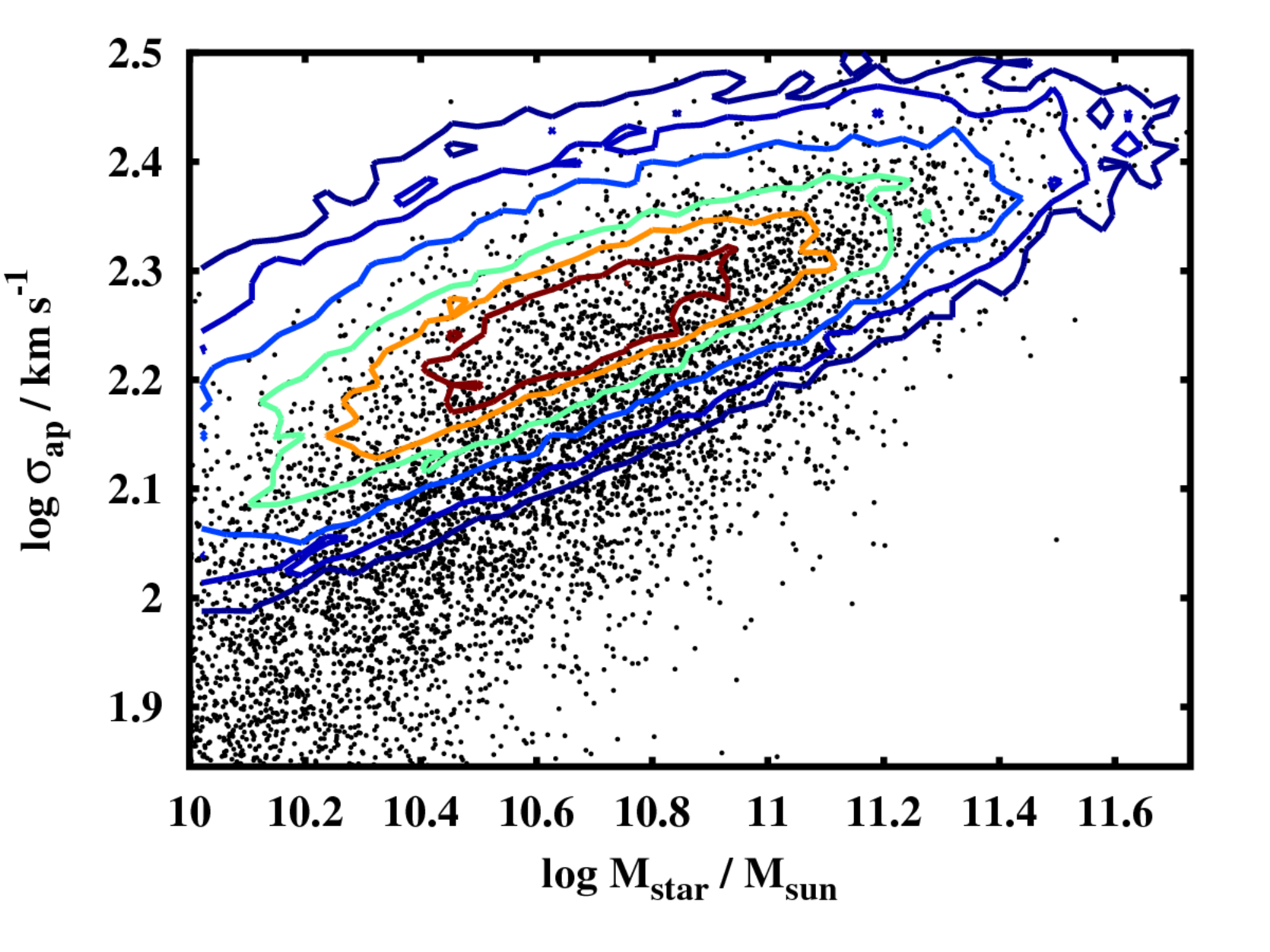}
    \label{fig:FJR_scatt}
  }
  \subfigure[$\nu \rightarrow -0.5$]
  {
    \includegraphics[width=0.395\textwidth]{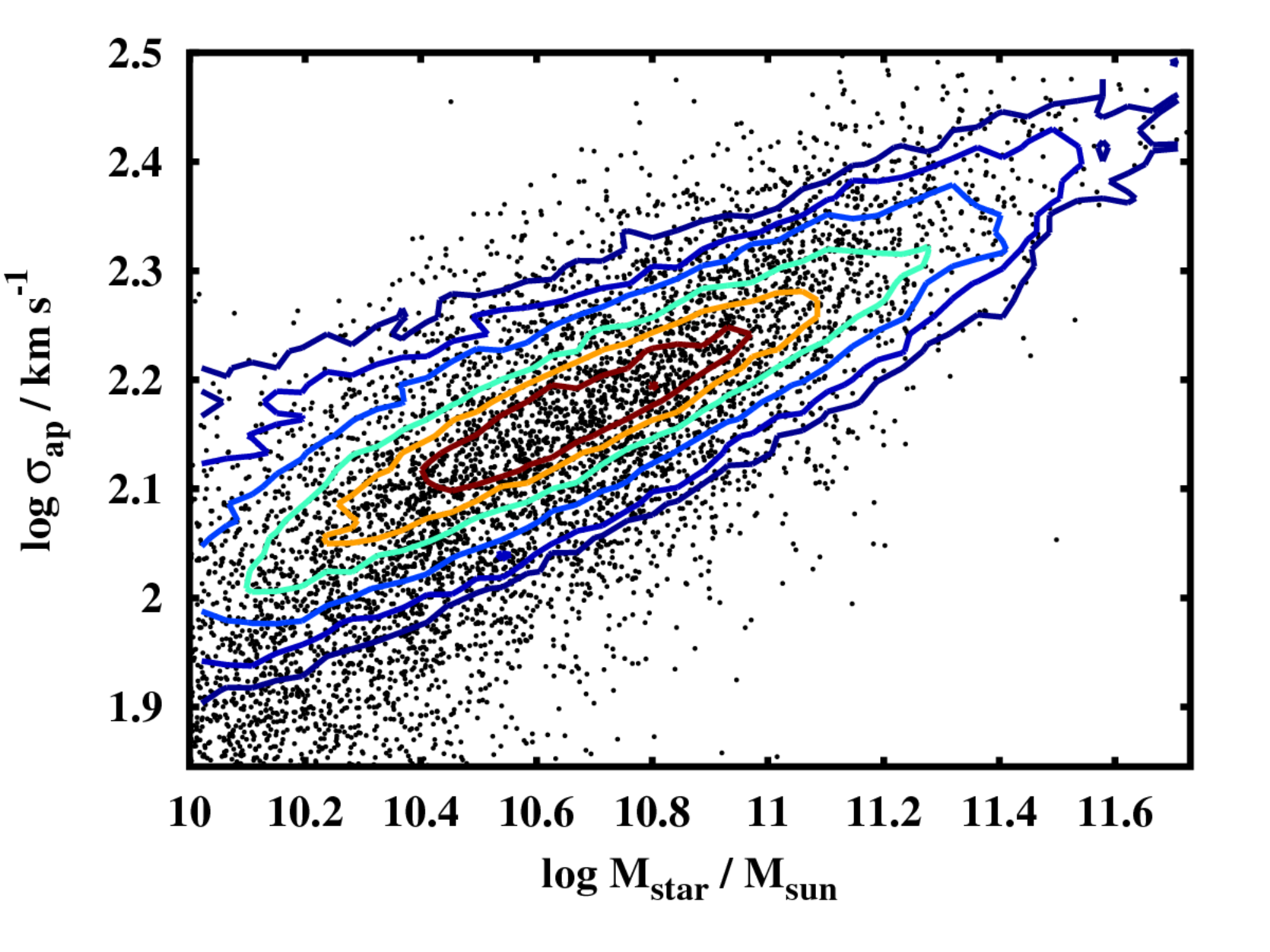}
    \label{fig:FJR_nu}
  }
  \subfigure[selection factor $\rightarrow \arctan(0.5)$]
  {
    \includegraphics[width=0.395\textwidth]{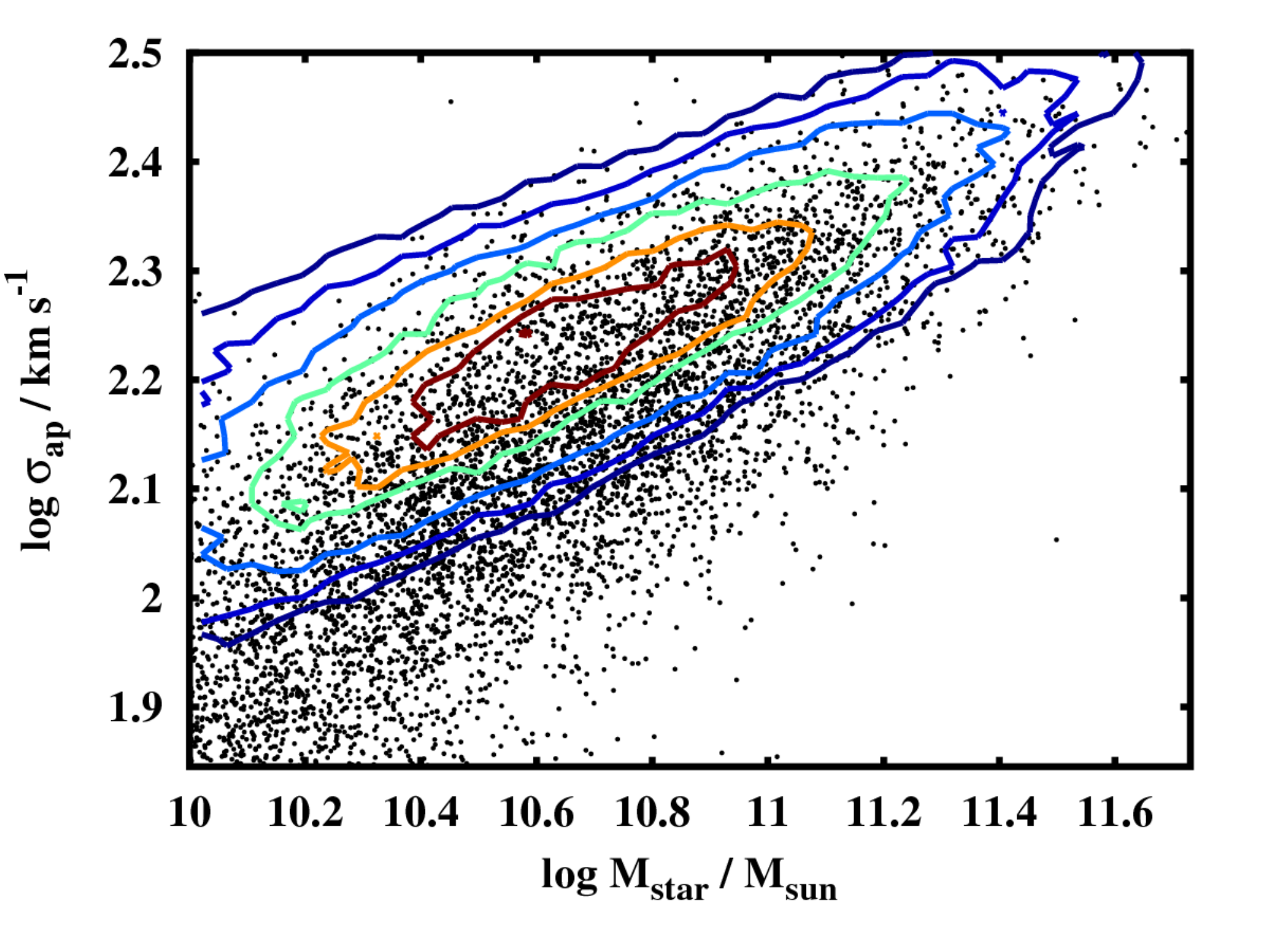}
    \label{fig:FJR_sel}
  }
  \subfigure[$\alpha=0.6$, AM scatter = 0.16, $\nu=-0.5$, selection factor=$\arctan(0.5)$]
  {
    \includegraphics[width=0.395\textwidth]{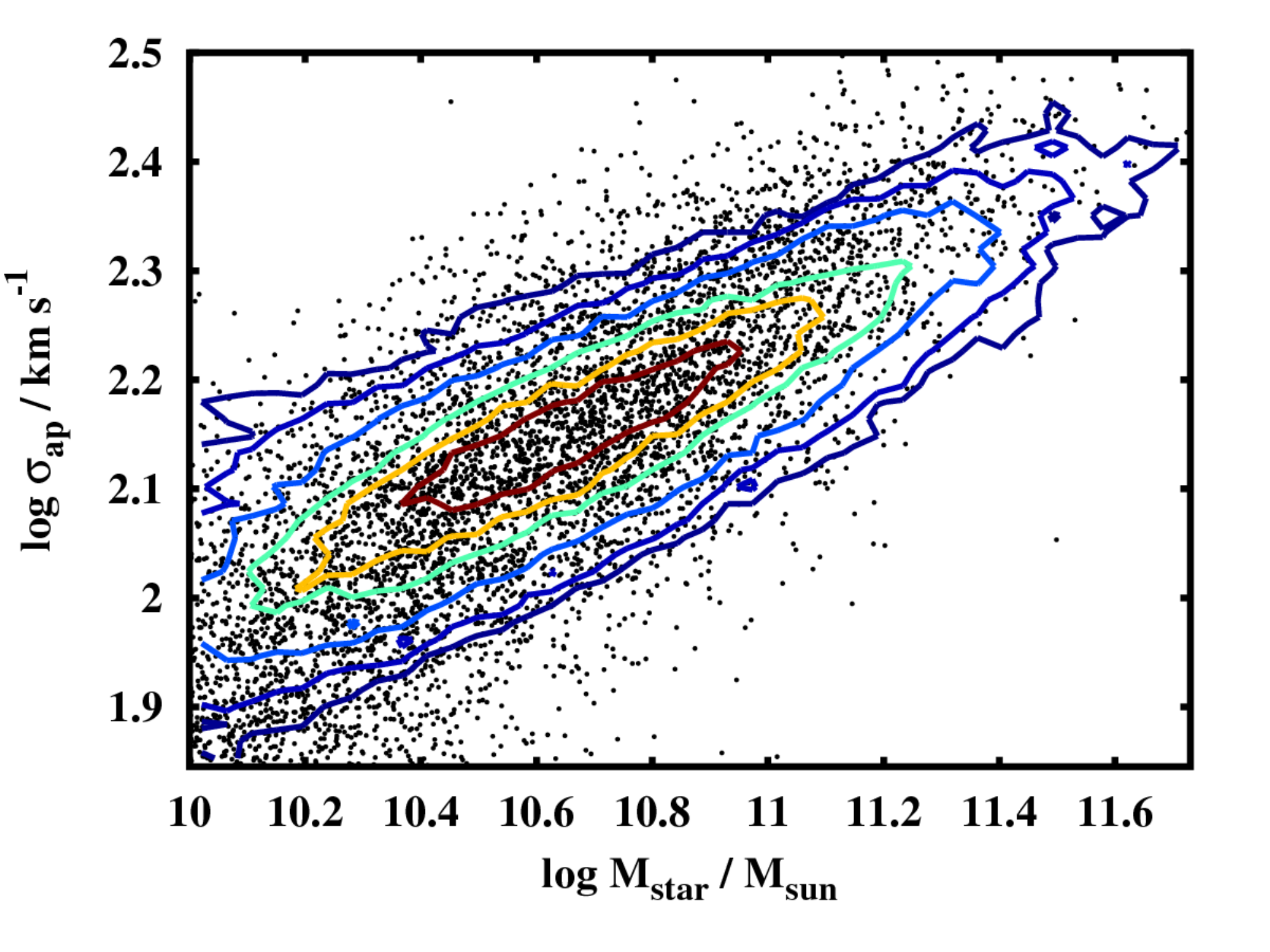}
    \label{fig:FJR_opt}
  }
  \subfigure[As Fig.~\ref{fig:FJR_opt}, but with $\beta \sim \mathcal{N}(0.3,0.35)$]
  {
    \includegraphics[width=0.395\textwidth]{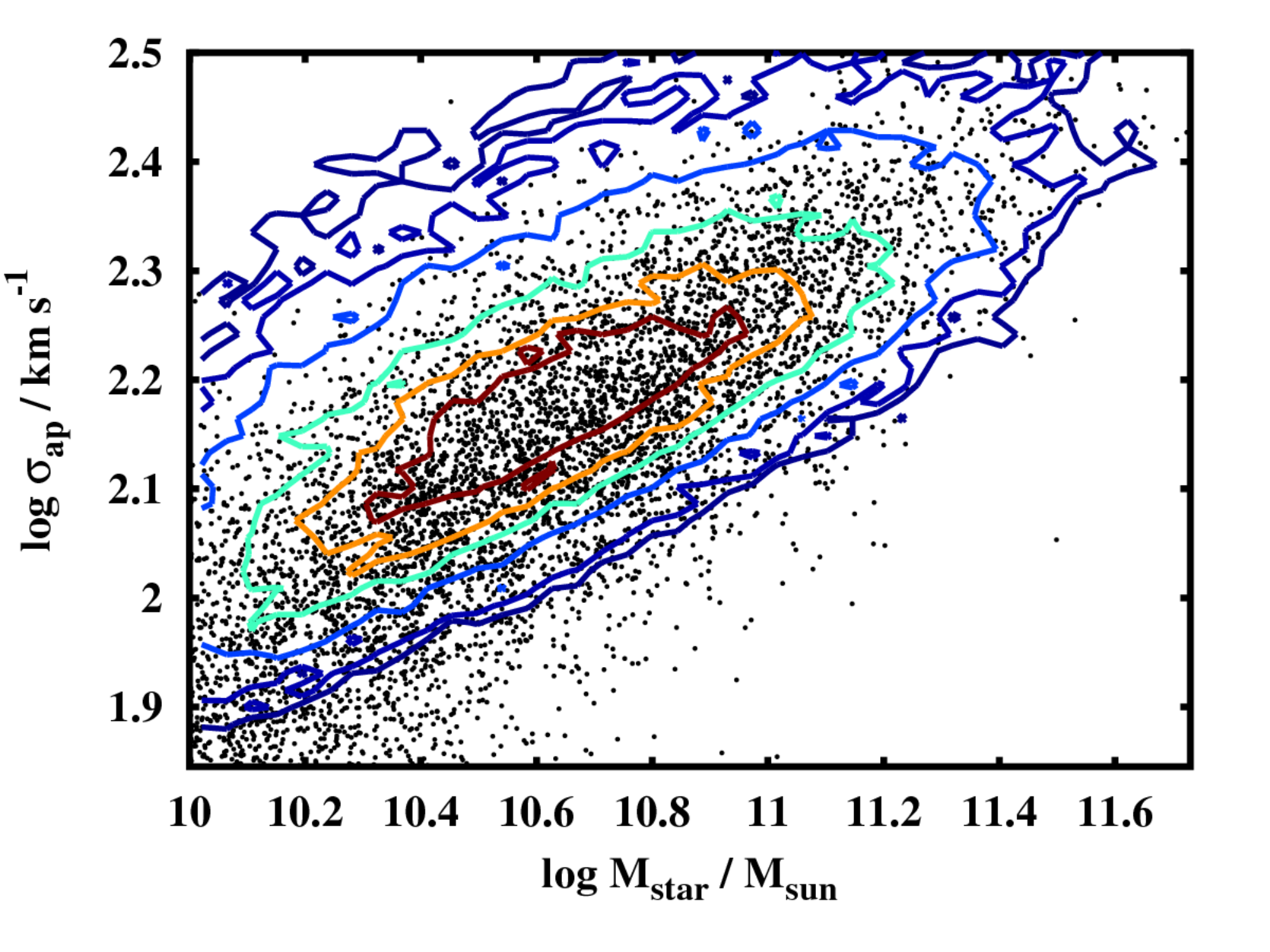}
    \label{fig:FJR_3beta}
  }
  \subfigure[As Fig.~\ref{fig:FJR_3beta}, but with $\langle \beta \rangle$ rising linearly from $-2.5$ at $\log(M_*/\mathrm{M_\odot})=10$ to its maximum possible value, $1$, at $\log(M_*/\mathrm{M_\odot})=11.7$]
  {
    \includegraphics[width=0.395\textwidth]{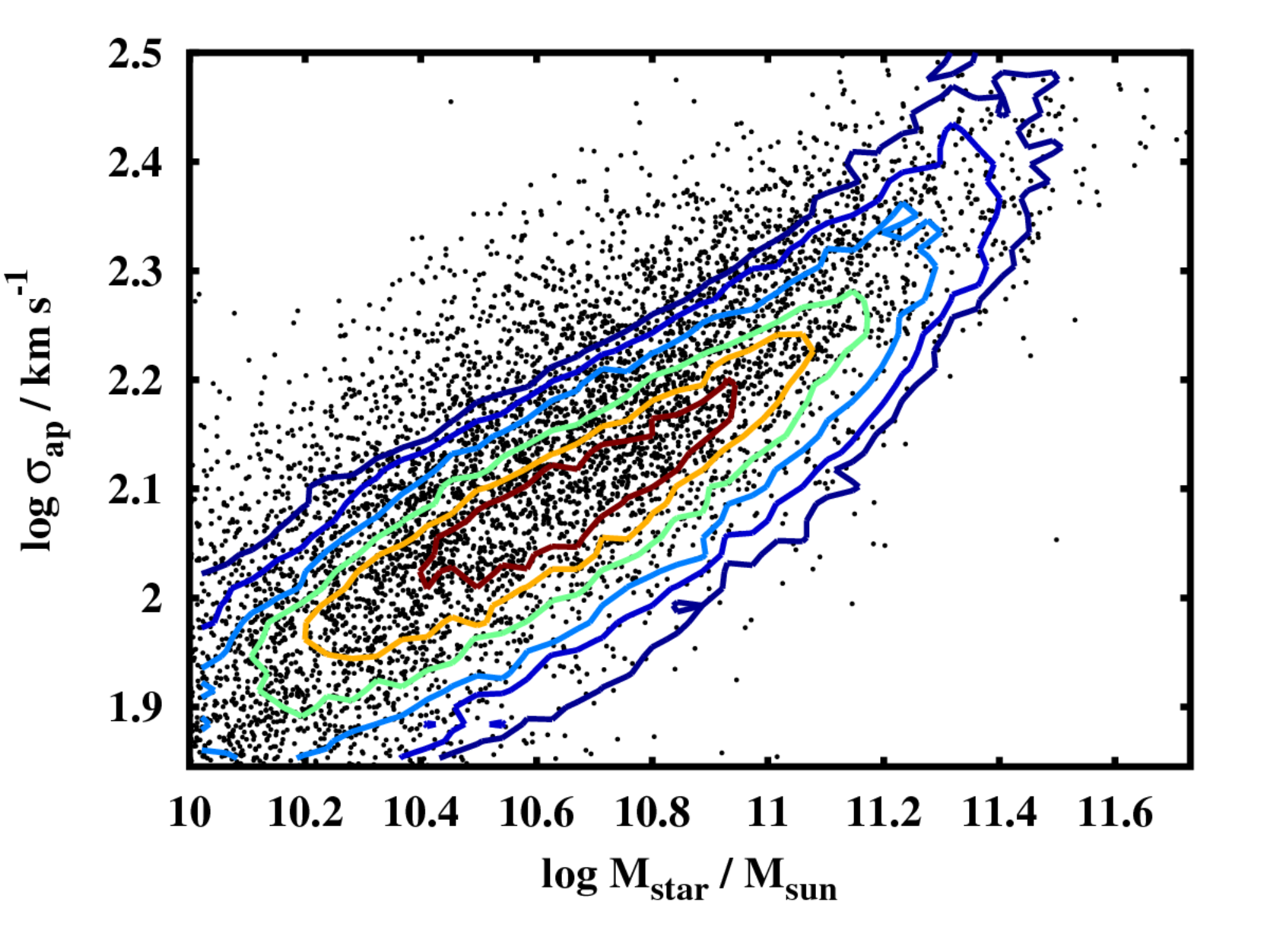}
    \label{fig:FJR_vbeta}
  }
  \caption{Contour plots showing the projections of the theoretical galaxy populations onto the FJR plane, compared to the NSA data in black. Each panel corresponds to a different choice for the model parameters, from the simplest no-DM case in the top left to the most complex AM model with mass-dependent anisotropy in the bottom right. The contour lines enclose 20, 40, 60, 80, 90 and 95 per cent of the model galaxies.}
  \label{fig:FJR_cont}
\end{figure*}

To generate a reference baryon-only prediction, we first consider the case of Newtonian gravity without dark matter haloes and with $\langle \beta \rangle = \sigma_\beta = 0$. The resulting FJR is displayed in Fig.~\ref{fig:FJR_noDM}, and the median values $\langle \boldsymbol{p} \rangle$ of the FJR and FP parameters across 1000 mock data sets drawn from the model are shown in the second row of Table~\ref{tab:table}. The standard deviations $\boldsymbol{\sigma}_p$ among the Monte Carlo realisations are found to be small, and their approximate fractional values (i.e. divided by the corresponding medians) are listed in Table~\ref{tab:table2}. According to the goodness-of-fit measure defined above, the likelihood of a model is formally low in terms of diagnostic $p_i$ if $\langle p_i \rangle$ differs from $p_{d,i}$ by more than a few digits in the third significant figure.

\begin{figure*}
  \subfigure[No-DM]
  {
    \includegraphics[width=0.395\textwidth]{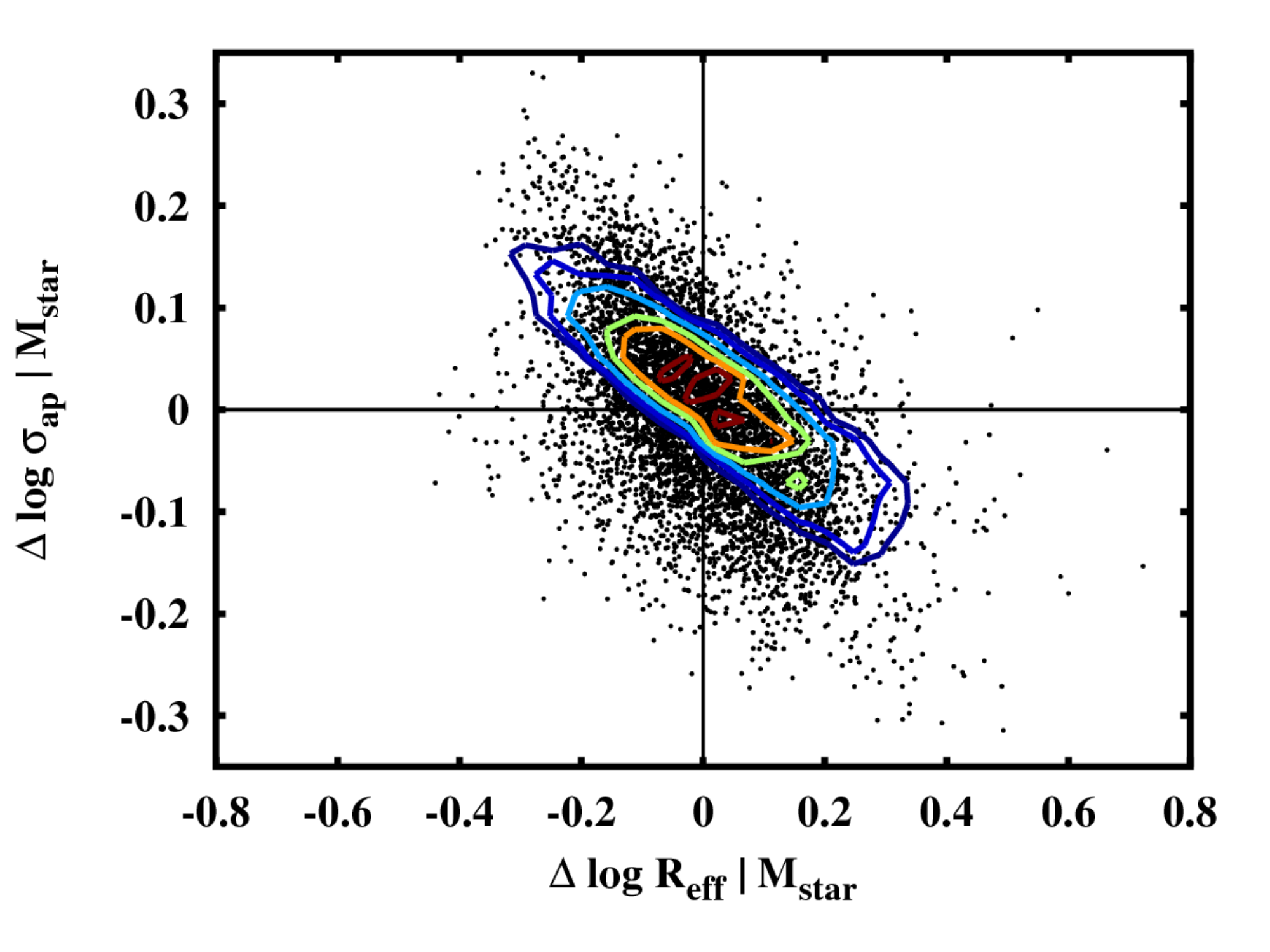}
    \label{fig:Rescorr_noDM}
  }
  \subfigure[$\alpha=0.6$, AM scatter = 0.16, $\nu=-0.5$, selection factor=$\arctan(0.5)$]
  {
    \includegraphics[width=0.395\textwidth]{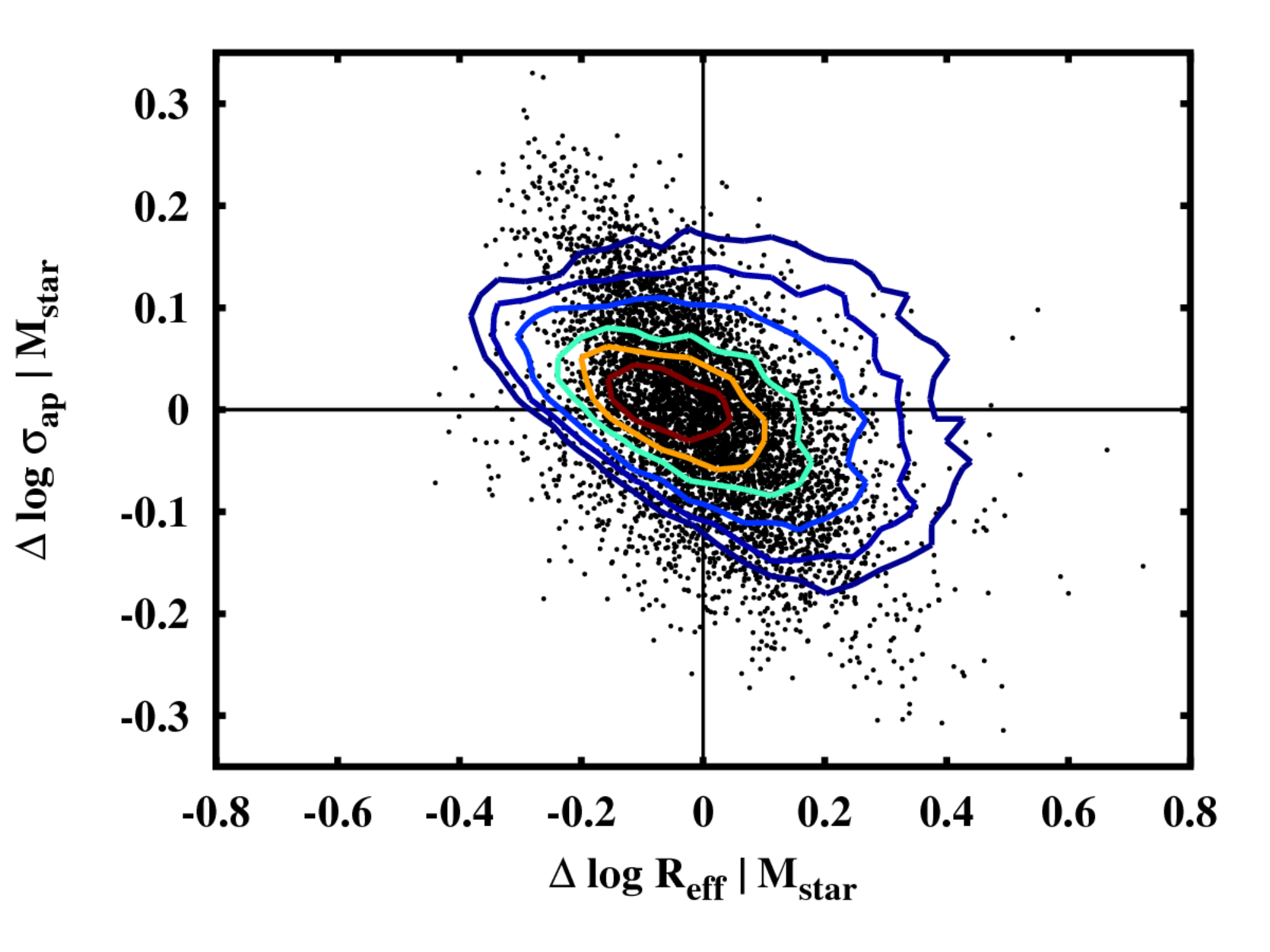}
    \label{fig:Rescorr_opt}
  }
  \subfigure[As Fig.~\ref{fig:Rescorr_opt}, but with $\beta \sim \mathcal{N}(0.3, 0.35)$]
  {
    \includegraphics[width=0.395\textwidth]{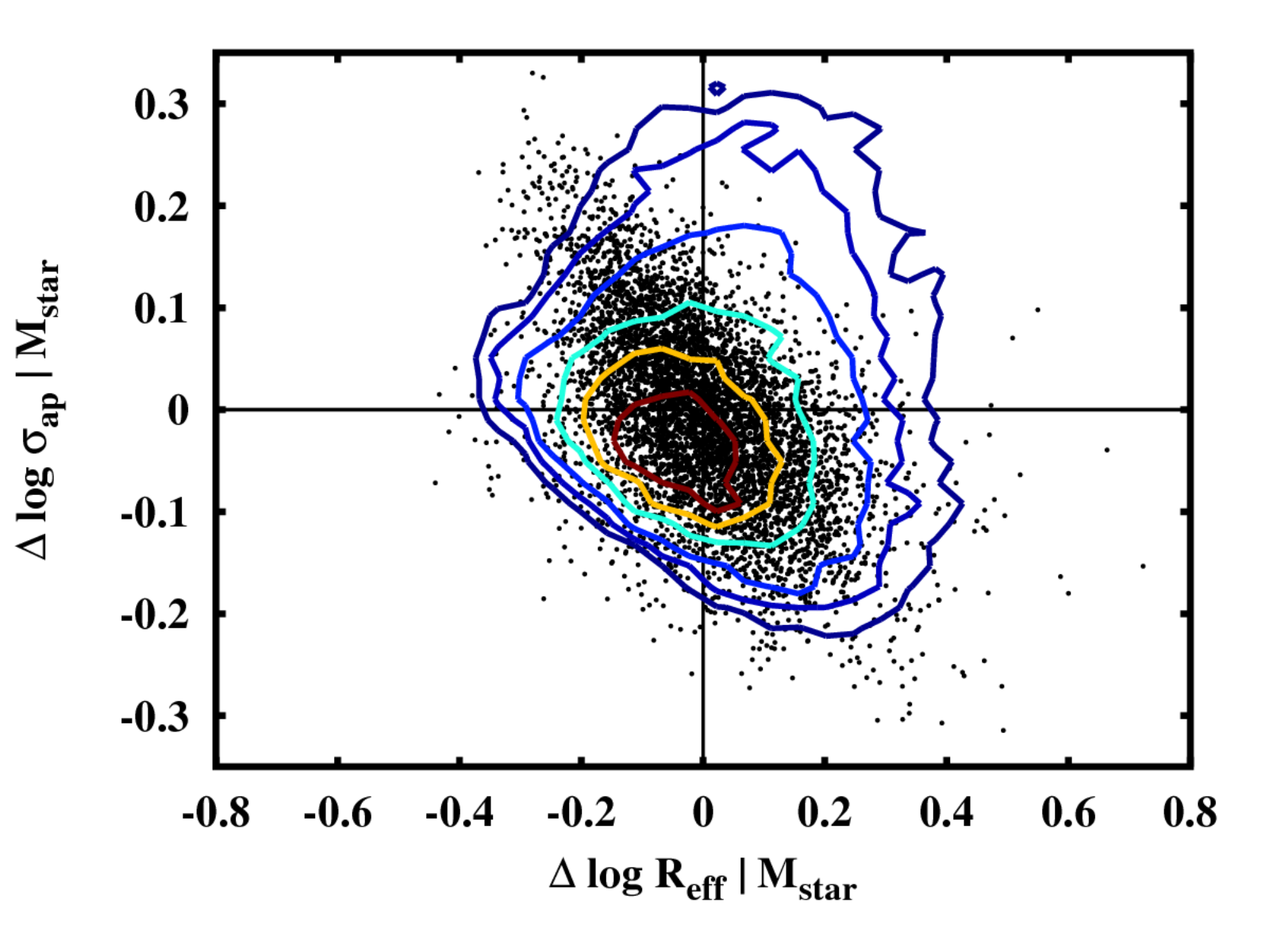}
    \label{fig:Rescorr_3beta}
  }
  \subfigure[As Fig.~\ref{fig:Rescorr_3beta}, but with $\langle \beta \rangle$ rising linearly from $-2.5$ at $\log(M_*/\mathrm{M_\odot})=10$ to its maximum possible value, $1$, at $\log(M_*/\mathrm{M_\odot})=11.7$]
  {
    \includegraphics[width=0.395\textwidth]{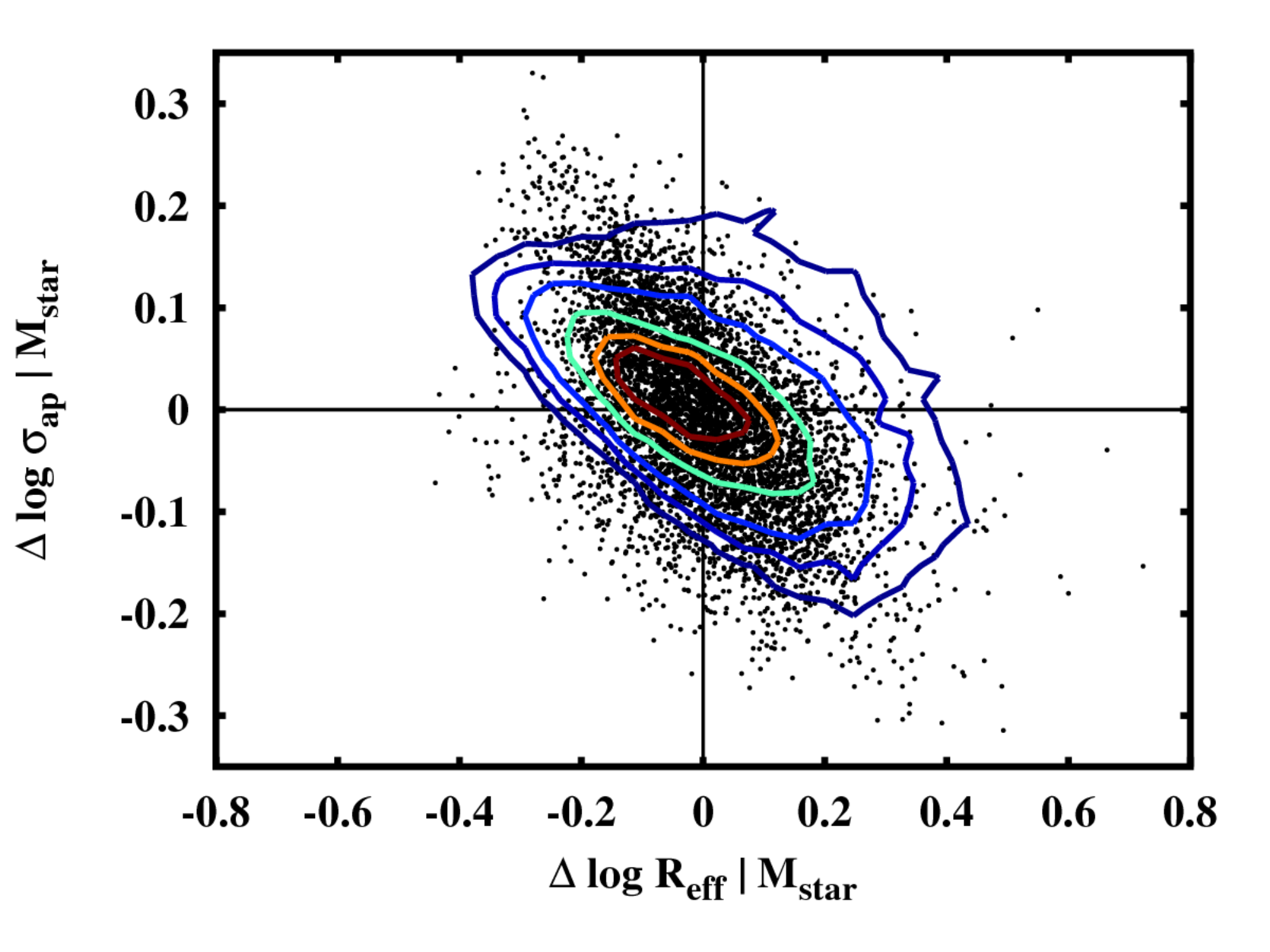}
    \label{fig:Rescorr_vbeta}
  }
  \caption{Correlation of the radius residuals of the MSR with the velocity residuals of the FJR, in the data (black points), and in 1000 mock data sets drawn from the models (contours). Each panel corresponds to different model assumptions, as labelled. The contour lines enclose 20, 40, 60, 80, 90 and 95 per cent of the model galaxies.}
  \label{fig:Rescorr_cont}
\end{figure*}

We see from Fig.~\ref{fig:FJR_noDM} that at low $M_*$ little dark matter is required to reproduce the observed velocity dispersion, while at high mass the predicted FJR lies significantly below that observed. Thus $m$ and $c_{FJR}$ are too low. As expected, $\kappa$ is very close to $-0.5$; the difference is due to variations among the galaxies in S\'{e}rsic index and the physical radius at which $\sigma_\mathrm{ap}$ is calculated. Both $s_{FJR}$ and $s_{FP}$ are too small, indicating that a significant source of scatter is absent. In addition, the scatter in $\log(M_*/\mathrm{M_\odot})$ is a factor of 5 smaller in the FP than the FJR: the inclusion of radius information is \emph{too} helpful in reducing the $\log(M_*)$ scatter in this case. In other words, mass and radius alone do not set the velocity as precisely as in this model; elliptical galaxies cannot be as simple as isotropic Newtonian S\'{e}rsic spheres. This is apparent also in Fig.~\ref{fig:Rescorr_noDM}, which correlates the residuals of mock and real galaxies from the best-fitting FJR and MSR. $\kappa$ is the slope of the best-fitting line in this space, and $s_{FP}$ measures the scatter of the points around this line.

\begin{table*}
  \begin{center}
    \begin{tabular}{l|c|c|c|c|c}
      \hline
      ($\alpha$, AM scatter, $\nu$, selection factor) &$m$ &$c_{FJR}$ &$s_{FJR}$ &$\kappa$ &$s_{FP}$\\ 
      \hline
      NSA data			& 0.313	& 2.13	& 0.095	& $-0.74$	& 0.127\\
      No-DM			& 0.253	& 2.03	& 0.065	& $-0.47$	& 0.055\\
      0.6, 0.16, 0.5, 1.4	& 0.220	& 2.25	& 0.067	& $-0.24$	& 0.205\\
      0.6, 0.4, 0.5, 1.4	& 0.186	& 2.24	& 0.078	& $-0.46$	& 0.221\\
      0.6, 0.16, $-0.5$, 1.4	& 0.229	& 2.16	& 0.063	& $-0.21$	& 0.194\\
      0.6, 0.16, 0.5, $\arctan(0.5)$	& 0.229	& 2.22	& 0.071	& $-0.35$	& 0.189\\
      0.6, 0.16, $-0.5$, $\arctan(0.5)$	& 0.238	& 2.12	& 0.067	& $-0.30$	& 0.178\\
      $\beta \sim \mathcal{N}(0.3,0.35)$	& 0.240	& 2.19	& 0.103	& $-0.79$	& 0.212\\
      $\beta \sim \mathcal{N}(2.06 \log(M_*/\mathrm{M_\odot}) - 23.1, 0.35)$	& 0.320	& 2.07	& 0.071	& $-0.33$	& 0.140\\
      \hline
    \end{tabular}
  \caption{Values for five key FJR and FP diagnostics in the data (first row), and their median values $\langle \boldsymbol{p} \rangle$ from 1000 mock data sets drawn from the model under various assumptions (remaining rows). $m$ is the FJR slope, $c_{FJR}$ its normalisation and $s_{FJR}$ its scatter, $\kappa \equiv \frac{d\sigma_\mathrm{ap}}{dR_\mathrm{eff}}|_{M_*}$, and $s_{FP}$ is the scatter around the best-fitting Fundamental Plane (see Eqs. 11 \& 12 and Table~\ref{tab:params2}). The numbers in the left-hand column indicate the values of $\alpha$, AM scatter, $\nu$ and selection factor, respectively, for the model in question. The final two rows use the same values for these parameters as the row above.}
  \label{tab:table}
  \end{center}
\end{table*}

\begin{table*}
  \begin{center}
    \begin{tabular}{l|c|c|c|c|c}
      \hline
      \% variation &$m$ &$c_{FJR}$ &$s_{FJR}$ &$\kappa$ &$s_{FP}$\\
      \hline
      Without DM		& 1	& 0.04	& 1	& 0.7	& 0.8\\
      With DM			& 1	& 0.04	& 2	& 5	& 2\\
      \hline
    \end{tabular}
  \caption{Approximate percentage variation $\boldsymbol{\sigma}_p/\langle \boldsymbol{p} \rangle \times 100$ per cent of the best-fitting FJR and FP diagnostic values $\boldsymbol{p}$ over 1000 Monte Carlo realisations. The eight cases in Table~\ref{tab:table} are reduced here to two: the results for all of the models that include DM (rows three to nine of Table~\ref{tab:table}) are approximately the same. A percent-level variation among realisations means that the likelihood of a model is formally low if the $\langle \boldsymbol{p} \rangle$ values it generates differ from $\boldsymbol{p}_d$ by more than a few digits in the third significant figure.}
  \label{tab:table2}
  \end{center}
\end{table*}

\subsection{AM model with $\beta=0$}
\label{sec:AM}

We now consider a fiducial AM model with $\alpha=0.6$, AM scatter = 0.16, $\nu=0.5$ and selection factor = 1.4 (Fig.~\ref{fig:FJR_fid} and Table~\ref{tab:table}, third row). These might be reasonable estimates a priori: the AM parameters are those preferred by~\citet{Lehmann}, the $\nu$ choice corresponds to moderate halo contraction, and the selection factor corresponds to a moderate preference for early-type galaxies to inhabit more massive or more concentrated haloes at fixed mass. The added dark matter, distributed in this way, has four main effects on the FJR and FP:

\begin{enumerate}

\item{} The normalisation of the FJR is increased due to the extra dynamical mass.

\item{} The velocity dispersions of low-mass galaxies are boosted more than those of high-mass galaxies, making the FJR shallower. This effect is exacerbated by the strong selection in this model: since the elliptical fraction falls towards lower stellar mass, the effect of selecting only the highest $\sigma_\mathrm{ap}$ objects to be ellipticals is greatest there.

\item{} The scatter of the FJR and (particularly) the FP are increased by the addition of halo variables ($M_\mathrm{vir}$ and concentration) into the effective formula for $\sigma_\mathrm{ap}$. These variables themselves have a spread of values at fixed $M_*$. The precision with which one can estimate $\log(M_*)$ from $R_\mathrm{eff}$ and $\sigma_\mathrm{ap}$ is only 1.5 times greater than from $\sigma_\mathrm{ap}$ alone due to the relative weakening of the correlation between the three observables.

\item{} $\kappa$ is increased, indicating that $\sigma_\mathrm{ap}$ varies less rapidly with $R_\mathrm{eff}$ at fixed $M_*$ than in the no-DM case. Now that $\sigma_\mathrm{ap}$ is set largely by the halo (and under our assumption that $R_\mathrm{eff}$ is independent of halo properties), the dependence of $\sigma_\mathrm{ap}$ on surface brightness is reduced.

\end{enumerate}

We now consider the effect of individual changes to the model parameters. We note first that varying $\alpha$ within reasonable bounds ($0 < \alpha < 1$;~\citealt{Lehmann}) has very little effect on any of our diagnostics. When $\alpha$ is larger, the rank of a halo in AM is conditioned more by its concentration than its virial mass, making the haloes of our high-$M_*$ galaxies more concentrated and less massive on average. However, the total dark matter mass within the aperture, which is responsible for setting $\sigma_\mathrm{ap}$, is only slightly larger at high $\alpha$. We note that the halo proxy has a significantly weaker effect on the FJR here than on the TFR in DW15. This is because in DW15 velocities were measured at larger radius, so the ratio of enclosed dark matter to baryonic mass was larger and the choice of halo proxy had a larger effect. In addition, while the squared TFR circular velocity is directly proportional to the enclosed dark matter mass, the FJR aperture velocity dispersion is determined by a set of three differential equations that depend non-trivially on density and cumulative mass. The result is a washing out of the subtle effect of the AM proxy.  We therefore fix $\alpha=0.6$ for the remainder of this work. 

The effects of variations in the other three model parameters are shown in rows four to six of Table~\ref{tab:table} and Figs.~\ref{fig:FJR_scatt}--\ref{fig:FJR_sel}, and may be accounted for as follows.

\begin{itemize}

\item{} The most significant effect of the AM scatter is on the scatter of the FJR and FP. Increasing the AM scatter increases the spread of halo properties assigned to galaxies of a given $M_*$, and hence the spread of the resulting velocity dispersions. A higher scatter also flattens the FJR; since the halo mass function is a falling function of mass, high AM scatter tends to associate lower mass haloes with galaxies at a given high $M_*$. Finally, increasing the AM scatter reduces $\kappa$.

\item{} $\nu$ largely sets the normalisation of the FJR: the lower it is, the more dark matter is expelled from the inner regions of a galaxy, and hence the lower $\sigma_\mathrm{ap}$. The normalisation of the FJR can be reconciled with that of the data by setting $\nu \approx -0.5$. In addition, lowering $\nu$ tends to reduce the rate at which $\sigma_\mathrm{ap}$ falls as $R_\mathrm{eff}$ rises, as manifest by an increase in $\kappa$. When $\nu>0$, more compact galaxies (i.e. lower $R_\mathrm{eff}$ at fixed $M_*$) pull more DM into the inner regions, boosting the rate at which $\sigma_\mathrm{ap}$ rises as $R_\mathrm{eff}$ falls. Conversely, when $\nu<0$ more compact galaxies \emph{expel} more DM from their inner regions, mitigating the rise in $\sigma_\mathrm{ap}$ due to the baryonic mass as $R_\mathrm{eff}$ falls.

\item{} A larger selection factor preferentially puts elliptical galaxies in more massive or more concentrated haloes at given $M_*$, increasing the normalisation of the FJR and reducing its scatter. Since the elliptical fraction falls towards low $M_*$, selection effects have largest impact in the low-mass regime, and hence increasing the selection factor lowers the FJR slope.

\end{itemize}

Table~\ref{tab:table} shows that the predicted $m$ and $c_{FJR}$ are brought into better agreement with the data by adopting $\nu = -0.5$ and selection factor = $\arctan(0.5)$. We display the results of such a model in Figs.~\ref{fig:FJR_opt} and~\ref{fig:Rescorr_opt}, and row seven of Table~\ref{tab:table}. We focus hereafter on the case \{$\alpha = 0.6$, AM scatter = 0.16, $\nu = -0.5$, selection factor = $\arctan(0.5)$\}.

\subsection{AM model with $\beta\neq0$}
\label{sec:AM_beta}

All models examined so far have assumed $\beta = 0$. We now consider two models where this constraint is relaxed, the first in which a $\beta$ value for each galaxy is drawn from a normal distribution with fixed mean $\langle \beta \rangle$ and standard deviation $\sigma_\beta$, and the second in which $\langle \beta \rangle$ is a function of $M_*$.

Detailed kinematic studies of ellipticals suggest that $\beta$ is typically fairly independent of radius and takes a value around 0.3 with an inter-galaxy scatter of roughly $0.1-0.4$~\citep{Gerhard_anisotropy, Thomas_anisotropy, Cappellari_anisotropy}. The results of a model with $\beta \sim \mathcal{N}(0.3, 0.35)$, separately for each mock galaxy, are shown in Fig.~\ref{fig:FJR_3beta}, Fig.~\ref{fig:Rescorr_3beta} and row eight of Table~\ref{tab:table}. This model agrees fairly well with the data in terms of both $s_{FJR}$ and $\kappa$, and the large effect of $\beta$ demonstrates its crucial importance in any study aimed at uncovering the theoretical significance of elliptical galaxy scaling relations. We distinguish and explain the effects of $\langle \beta \rangle$ and $\sigma_\beta$ as follows.

\begin{enumerate}

\item{} All else being equal, the larger a galaxy's $\beta$, the greater its $\sigma_\mathrm{ap}$. This is because larger $\beta$ corresponds to more radial stellar motions, which provide a larger contribution to $\sigma_r$ and hence $\sigma_\mathrm{los}$ (Eqs. 6--9). Thus $c_{FJR}$ rises with $\langle \beta \rangle$.

\item{} A larger $\beta$ increases $\sigma_\mathrm{ap}$ for smaller galaxies more than it does for larger ones. Increasing $\langle \beta \rangle$ therefore increases the average $\sigma_\mathrm{ap}$ of galaxies at fixed $M_*$ and low $R_\mathrm{eff}$ more than those at the same $M_*$ but high $R_\mathrm{eff}$, reducing $\kappa$.

\item{} At small $R_\mathrm{eff}$, a large $\beta$ increases $\sigma_\mathrm{ap}$ more than a small $\beta$ decreases it, and vice versa at large $R_\mathrm{eff}$. Hence increasing $\sigma_\beta$ at fixed $\langle \beta \rangle$ increases the average $\sigma_\mathrm{ap}$ of galaxies at fixed $M_*$ and low $R_\mathrm{eff}$, while decreasing it at high $R_\mathrm{eff}$. The result is a decrease in $\kappa$.

\item{} Since $\beta$ is assumed to be uncorrelated with galaxy or halo properties, the effect of $\sigma_\beta$ roughly sums in quadrature with the other sources of scatter to increase $s_{FJR}$ and $s_{FP}$.

\end{enumerate}

We now consider a simple toy model in which $\langle \beta \rangle$ depends on $M_*$. This is motivated by the observation from Fig.~\ref{fig:FJR_cont} and Table~\ref{tab:table} that none of the model refinements made so far have had a significant effect on the \emph{slope} of the FJR, which has remained in all cases too low. We therefore show in Figs.~\ref{fig:FJR_vbeta} and~\ref{fig:Rescorr_vbeta}, and row nine of Table~\ref{tab:table}, the effect of the alternative model $\beta \sim \mathcal{N}(2.06 \log(M_*/\mathrm{M_\odot}) - 23.1, 0.35)$, in which $\langle \beta \rangle$ rises from $-2.5$ at $M_*=10^{10} \mathrm{M_\odot}$ to 1.0 at $M_*=10^{11.7} \mathrm{M_\odot}$.\footnote{Although we do not intend to motivate this model on prior grounds, a weak trend in this direction may be expected: the aperture size is fixed at 1.5'', more massive galaxies tend to have larger $R_\mathrm{eff}$, and $\beta$ tends to fall slowly with $r/R_\mathrm{eff}$ beyond a certain point (e.g.~\citealt{Gerhard_anisotropy}).} The FJR slope in this model is increased to that in the data, and furthermore $s_{FP}$ is reduced. However, the model FJR has acquired a curvature not present in the data, and $\kappa$ is significantly increased. Within our parameter space, the FJR curvature could be reduced only by setting $\langle \beta \rangle$ even lower at $M_*=10^{10} \mathrm{M_\odot}$ ($\beta$ can never exceed 1), which would further increase $\kappa$ and necessitate an increase in $\nu$ to mitigate the fall in $c_{FJR}$. We therefore conclude that introducing a dependence of $\langle \beta \rangle$ on $M_*$ does not offer a particularly attractive prospect for further improving FJR and FP predictions.

Note that a positive $\langle \beta \rangle$ and $\sigma_\beta$ would increase the predicted FJR intercept and scatter also in the no-DM case, bringing it into better agreement with the data, and including a positive $\langle \beta \rangle$ dependence on $M_*$ would increase the slope. Within the broader parameter space opened up by the anisotropy, it is not clear therefore that the no-DM model can be excluded by the FJR alone.

\subsection{Degeneracies}
\label{sec:degeneracies}

Now that all the parameters of our model have been introduced and their effects described, we discuss their degeneracies. The parameters $\nu$, $\langle \beta \rangle$ and the selection factor have a degenerate effect on the normalisation of the FJR. There is also a strong degeneracy between the AM scatter and $\sigma_\beta$, both of which predominantly affect $s_{FJR}$ and $s_{FP}$. However, strong prior constraints exist on the AM scatter~\citep[e.g.][and references therein]{Lehmann}, so we have chosen to fix its value here. We note however, that although most current measurements are consistent with a constant value or weak trend of AM scatter with mass, the constraints weaken significantly at low mass. Finally, the selection factor, AM scatter and effective dependence of $\beta$ on $M_*$ are degenerate in their effect on $m$, and $\nu$, $\sigma_\beta$ and the AM scatter are degenerate in determining $\kappa$. Alternative sets of parameter values capable of adequately reproducing the observed $c_{FJR}$, $s_{FJR}$ and $\kappa$ are therefore likely to exist.

\section{Discussion}
\label{sec:discussion}

We have investigated a series of AM-based models, which have previously been shown to match several spatial statistics of galaxies, in order to assess the extent to which important characteristics of the Faber--Jackson relation and Fundamental Plane are expected in contemporary galaxy formation scenarios. While some aspects of the relations (the normalisation and scatter of the FJR, and the rate of change of velocity with size at fixed mass) may be satisfactorily accounted for, others (the FJR slope and FP scatter) are challenging to match with our approach. In Section~\ref{sec:contraction} we discuss our preference for halo expansion in the context of previous findings in the literature, in Section~\ref{sec:selection} we comment on the possibility that early-type galaxies tend to inhabit more massive or more concentrated haloes at given stellar mass, and in Section~\ref{sec:discrepancies} we elaborate on the discrepancy in the FP scatter and consider potential solutions.

\subsection{FJR normalisation, halo contraction, and the choice of IMF}
\label{sec:contraction}

In order to obtain the correct magnitude of the aperture velocity dispersions, we require moderate expansion of haloes in response to disc formation ($\nu \approx -0.5$). This roughly agrees with the TFR calculations in DW15, which favour $\nu \approx 0$ for mass-based AM and $\nu \approx -0.5$ for the more realistic case of velocity-based matching. These results may be considered an extension of the cusp--core~\citep{cusp-core} and Too Big To Fail~\citep*{TBTF} problems to higher mass: the NFW profile with a canonical mass-concentration relation puts too much dark matter in the central regions of $M_* = 10^{10} - 10^{11.7} \mathrm{M_\odot}$ galaxies.

The necessity of halo expansion has already been noted by several observational and theoretical studies of early-type galaxies.~\citet{Borriello} find that reproducing the low normalisation, scatter and curvature of the FP requires either that the haloes surrounding ellipticals have significantly lower mass or concentration than predicted by $N$-body simulations, or that they are substantially cored.~\citet{Romanowsky} observe Keplerian fall-off of the velocity dispersion profiles of three bright ellipticals, indicating a low dark matter fraction out to a significant radius. Indeed, many studies indicate that dark matter constitutes only a small fraction of the total mass in the central regions of ellipticals (e.g.~\citealt{Salucci, Mamon_Lokas, Tiret_Combes_MOND_VD, Napolitano, Cappellari_2, Cappellari, Nigoche}). Finally, since the majority of ellipticals under investigation here are high surface brightness systems, our findings add to the growing body of evidence that higher stellar surface mass density implies lower enclosed dark matter mass~\citep{Sanders_MDA, McGaugh_MDA, Gentile, Tiret_Combes, Sonnenfeld}. Indeed, it is found from the slope of the NSA MSR that the surface mass density at $R_\mathrm{eff}$ falls on average as $M_*$ increases, while Fig.~\ref{fig:FJR_noDM} demonstrates that the inferred dark matter mass rises.

While ``bottom up'' studies prioritising detailed treatment of the data over cosmological predictions typically find evidence for low or negligible dark matter fractions in the central regions, several ``top down'' studies that start from a cosmological model argue that halo \emph{contraction} is required in order to fit the FJR normalisation. We discuss two.

\citet{TG} compile a set of dynamical mass measurements at 10 kpc from the centres of elliptical galaxies, and combine this with TFR observations to create a single luminosity--velocity data set. They find that elliptical galaxies tend to have larger circular velocities than spirals at fixed luminosity. Taking the theoretical prediction to be the halo plus baryon velocities at 10 kpc after AM, they argue that halo contraction is necessary to reproduce the observations. Comparing the measured NSA FJR presented here to standard Tully--Fisher relations (e.g. that of~\citealt{P07}), we on the other hand find the FJR to have a lower normalisation and higher slope. Only at the very largest stellar masses ($\sim 10^{11.5} \mathrm{M_\odot}$) do the elliptical galaxy aperture velocity dispersions exceed the spiral galaxy $V_{80}$ values.

\citet{Dutton_1} combine a number of empirical power-law correlations relating galaxy size, gas content, halo virial mass, and halo concentration with stellar mass to generate a mock TFR and FJR. They do not in fact solve the Jeans equation to model $\sigma_\mathrm{ap}$, but instead adopt an average empirical relation between the circular velocity at $R_\mathrm{eff}$ (which is output by their model) and the velocity dispersion there, and an additional relation between that and the integrated velocity dispersion within the SDSS aperture. Their fiducial model with halo contraction generates an FJR with slope and intercept in almost perfect agreement with their data, i.e. with a considerably lower normalisation and higher slope than ours.

The differences between our results and those of~\citet{Dutton_1} may be largely attributable to the use of different stellar-to-halo mass ratios: where~\citet{Dutton_1} employs a compilation of satellite kinematics and weak lensing observations, we use abundance matching. By direct comparison, we find that our halo masses are around $0.2-0.3$ dex higher at fixed $M_*$, resulting in an increased need to move dark matter out of the central regions in order to match the FJR normalisation. These differences are within the uncertainties of direct measurements of the stellar-to-halo mass ratio (e.g.~\citealt{Uitert}), but have an important effect of mass modelling.

The choice of IMF is another important systematic. The Chabrier IMF that we use is among the lightest of those considered routinely in the literature for early-type galaxies; adopting the heavier Kroupa or Salpeter IMF would increase stellar masses and hence reduce the inferred dark matter mass. IMF studies in the literature may be settling on a non-universal IMF for early-type galaxies which increases roughly linearly from Chabrier at $\log(\sigma_e/\mathrm{km} \: \mathrm{s}^{-1}) \approx 1.8$ to Salpeter at $\log(\sigma_e/\mathrm{km} \: \mathrm{s}^{-1}) \approx 2.3$ (\citealt{Dutton_13};~\citealt{Cappellari_review} and references therein). Significantly, it is found not only that this IMF favours low central dark matter fractions ($\sim 13$ per cent), but also that the stellar mass Fundamental Plane is then fully consistent with the virial plane $M_* \propto \sigma^2 R$, with zero scatter. This suggest that were we to adopt this IMF in our analysis we would find the data to be in agreement with our no-DM prediction; in particular, the $m$ and $s_{FP}$ values measured from the data would be significantly reduced, and $\kappa$ increased.

Changing the IMF would also affect the FJR and FP output by our model, since greater stellar masses at the bright end correspond to larger stellar mass fractions with the halo mass function fixed. However, predicted velocities would increase along with the stellar masses, causing a shift roughly along the FJR and hence impacting the predicted relations less than the observed ones. The non-universal IMF of~\citet{Cappellari_review} would therefore still require $\nu < 0$. More generally, while the choice of IMF may affect any of the FJR or FP diagnostics in Table~\ref{tab:table}, the true IMF is very unlikely to generate a higher FP scatter in the data. Hence the $s_{FP}$ discrepancy could not be eliminated in this way.

Additional sources of discrepancy between our results and those of~\citet{TG} and~\citet{Dutton_1} may include a potential systematic error associated with combining incommensurate $\sigma_\mathrm{ap}$ and $V_\mathrm{rot}$ measurements (measured at different radii and governed by qualitatively different equations), and an offset between data and model velocity dispersions when the latter are not derived from Jeans analysis. We stress the importance of measuring velocities in precisely the same way in real and mock galaxies. Finally, we note that~\citet{Dutton_1} assume galaxy and halo variables to be correlated with $M_*$ only, while we include also their expected and observed correlations with one another. There is no guarantee that these covariances are insignificant.

\citet{Dutton_1} argue that simultaneously matching the normalisation of the TFR and FJR requires that early-type galaxies either experience stronger halo contraction in response to disc formation than late-types, or have a heavier IMF. We note that the relative amount of halo contraction or expansion required by spirals and ellipticals depends on whether the two types of galaxies are placed in haloes of systematically different mass or concentration. This effect is included in our framework via the selection factor. The higher the selection factor, the stronger the expansion required for ellipticals relative to spirals. In particular, we find that a high selection factor (i.e. the haloes of more elliptical galaxies have more mass in the central regions at fixed $M_*$) favours $\nu \approx -0.4$ for spirals (DW15), and $\nu \approx -0.6$ for ellipticals (see Fig.~\ref{fig:FJR_nu} and row five of Table~\ref{tab:table}), while random selection (selection factor = arctan(0.5)) favours $\nu \approx -0.6$ for spirals and $\nu \approx -0.4$ for ellipticals. Selection effects are important in any attempt to model the dynamics of spiral and elliptical galaxies within the same framework. 

Our model assumes that the dependence of the halo expansion factor on $M_*$ is the same as that of adiabatic contraction, viz., maximised at $M_* \approx 10^{10} \mathrm{M_\odot}$ where $M_*/M_\mathrm{vir}$ is largest. While this dependence is motivated in the case of contraction ($\nu > 0$), there is no guarantee that it would be preserved by the baryonic feedback processes presumed to cause halo expansion in the case $\nu < 0$. We see from Figs.~\ref{fig:FJR_opt} and~\ref{fig:FJR_3beta} that to bring the predicted FJR slope into better agreement with the observations would require even more expansion at $M_* \approx 10^{10} \mathrm{M_\odot}$ relative to $M_* \approx 10^{11.5} \mathrm{M_\odot}$ where the baryon fraction is very low. Expansion schemes which act to flatten cusps at intermediate $M_*/M_\mathrm{vir}$ (e.g.~\citealt{diCintio}) would not help here. The sensitivity of our halo response to $M_*/M_\mathrm{vir}$ makes our constraint on $\nu$ dependent on our being able to probe lower stellar masses; for the very highest mass galaxies some halo contraction may be permitted.

Two potential alternatives to halo expansion would require only slight deviation from our assumptions. One possibility is that $\Omega_\mathrm{m}$ or $\sigma_8$ is too large, resulting in haloes of too high mass or concentration being generated by the $N$-body simulation. However, we showed in Appendix B of DW15 that a significant effect on galaxy velocities would require these variables to change by more than their cosmological uncertainties (a conclusion also reached by~\citealt{Dutton_1}). Alternatively, the AM model could be changed to modify the relation between stellar mass and halo mass and concentration. In particular, lowering the normalisation of the FJR would require galaxies in the range $10^{10} \mathrm{M_\odot} < M_* < 10^{11.7} \mathrm{M_\odot}$ to lie in less massive or less concentrated haloes, and steepening the relation would require this effect to be larger at $10^{10} \mathrm{M_\odot}$ than $10^{11.7} \mathrm{M_\odot}$. However, this effect would have to be large to affect $c_{FJR}$ and $m$ significantly: it can be seen from Fig.~\ref{fig:FJR_noDM} there is little need for dark matter at $M_* \approx 10^{10} \mathrm{M_\odot}$, and hence $M_\mathrm{vir}$ would have to be lowered to near $M_*$ to obtain agreement here. At $M_* \approx 10^{11.5} \mathrm{M_\odot}$, we estimate that $M_\mathrm{vir}$ would have to be increased by a factor of $\sim 4$, or halo concentrations by $\sim 30$ per cent, to increase $\sigma_\mathrm{ap}$ by 0.1 dex.

\citet{Dutton_13} construct a framework in which to constrain the stellar mass dependence of the IMF and halo response to disk formation using the slope and normalisation of the FJR and the tilt of the FP (our $m$, $c_{FJR}$ and $\kappa$ statistics). In agreement with our analysis, they find that a universal value of the IMF and $\nu$ underpredicts the FJR slope, and that a baryon-only (mass follows light) model with IMF slightly heavier than Chabrier best accounts for $\sigma_\mathrm{ap}$ at low $M_*$. Halo contraction is disfavoured for $\log(M_*/M_\odot) \lesssim 11.4$, but may be allowed at higher mass provided the IMF is sufficiently light. The value they derive for the tilt of the observed FP is dissimilar to ours, and in particular lies on the other side of the virial value $\kappa = -0.5$. This may be due to different sample selection (our cuts on S\'{e}rsic index and light concentration are replaced by one on colour, and they go out to a higher redshift $z=0.3$), different baryonic mass modelling (they use a de Vaucouleurs plus exponential fit where we take a general S\'{e}rsic index $n$), or different stellar mass determinations. Nevertheless we agree that the strong dependence of $\sigma_\mathrm{ap}$ on halo variables induced by contracted or NFW haloes washes out the FP tilt to an unacceptable level, and hence overpredicts $\kappa$. We note that while indeed the average radial orbit anisotropy $\langle \beta \rangle$ has only a small effect on $c_{FJR}$ and $\kappa$ within the range of its observed values, the variation among galaxies, $\sigma(\beta)$, is less well constrained a priori and may have a larger effect.

We note finally that further insight into the shape of dark matter density profiles in this stellar mass range may come from combining precise probes of the inner regions (e.g. aperture velocity dispersion) with measurements of the large-scale mass distribution from weak lensing. While some analyses of this type have already been performed (e.g.~\citealt{Schulz}), these studies can be significantly expanded and improved with the current generation of large photometric surveys such as DES and HSC.

\subsection{Morphology selection effects}
\label{sec:selection}

We have found that lowering the selection factor -- that is, reducing the relative preference for ellipticals to live in more massive or more concentrated haloes at fixed $M_*$ -- improves the model prediction of the FJR slope, intercept and scatter, as well as the average rate at which $\sigma_\mathrm{ap}$ varies with $R_\mathrm{eff}$ at fixed $M_*$. The opposite was argued for by DW15 from the normalisation of the TFR. Taken in combination, these results suggest a scenario in which the overall dark matter mass in \emph{all} galaxies is reduced (e.g. by halo expansion), rather than one in which the two types of galaxy populate haloes with markedly different dynamical properties. It does however remain possible to situate early-types in more massive haloes by increasing the degree of their halo expansion. A natural next step would be to model these two populations and their selection simultaneously in a fully self-consistent way.

\subsection{The scatter of the Fundamental Plane}
\label{sec:discrepancies}

We have found two of our five FJR and FP diagnostics intractable within our framework: the FJR slope $m$ and the FP scatter $s_{FP}$. While the slope prediction may be improved by a more sophisticated model for radial orbit anisotropy $\beta$, a more realistic IMF, or a change to the mass-dependence of halo expansion, the discrepancy in the FP scatter likely points to a deeper problem with the model. We discuss this issue here.

A number of authors have suggested previously that the scatter in the FP predicted by $\Lambda$CDM is too large (e.g.~\citealt{Sanders_FP, Borriello, Salucci}); our contribution is to quantify these claims precisely within the AM framework. When galaxies are situated in haloes, scatter in halo mass and concentration at fixed $\sigma_\mathrm{ap}$ and $R_\mathrm{eff}$ provide additional scatter in $M_*$: roughly, $\Delta{\log(M_*)}$ is increased by $\Delta{\log(M_h(R_\mathrm{ap}))} \approx 0.2 \; \mathrm{dex}$ (see Table~\ref{tab:table}). Further, in all of our models that include dark matter (rows three to nine of Table~\ref{tab:table}), the reduction in scatter in progressing from the FJR to the FP is much lower than in the data. In the presence of a halo, the relative dependence of $\sigma_\mathrm{ap}$ on $R_\mathrm{eff}$ is reduced, and hence so too is the increase in the precision with which velocity may be determined when its mean dependence on size is accounted for in the optimum way.

We have assumed that galaxy size is correlated only with stellar mass and S\'{e}rsic index, and not separately with halo properties at a given stellar mass and profile. Thus varying $R_\mathrm{eff}$ at fixed $M_*$ (and $R_\mathrm{ap}$) affects only the baryonic mass within the aperture and leaves the dark matter mass unchanged. The failure of our models to match the observed FP scatter while at the same time performing well in terms of $s_{FJR}$ and $\kappa$ suggests a correlation between galaxy size and halo properties that we have not included. We note for example that~\cite{Kravtsov13} have shown that galaxy half-mass radius is tightly correlated with halo mass, though it is unclear from that work whether this is true at fixed stellar mass.  One such possibility for an underlying correlation between size and halo properties was explored in DW15, where size was determined by the assumption that galaxy and halo specific angular momentum are proportional. This correlation acts to increase the rate at which $V$ falls as $R$ rises, since larger galaxies end up in less concentrated haloes at fixed $M_*$ and halo spin. A model along these lines would therefore likely reduce $s_{FP}$ (since the dependence of $\sigma_\mathrm{ap}$ on $R_\mathrm{eff}$ would be strengthened relative to its dependence on halo properties that do not appear in the FP), and decrease $\kappa$. Since predicted and observed $\kappa$ values agree under the empirical size model implemented here, this would run the risk of making $\kappa$ too negative. This could perhaps be remedied by reducing $\sigma_\beta$, but at the price of predicting too low an FJR scatter. Furthermore, this type of correlation between $R_\mathrm{eff}$ and halo spin was shown in DW15 to overpredict the scatter in size at fixed $M_*$.

An alternative approach would be to start with the observed values of $R_\mathrm{eff}$, as we have done here, but correlate $R_\mathrm{eff}$ explicitly with a halo property such as spin or concentration at fixed $M_*$. This is a form of conditional abundance matching~\citep{CAM}, and would retain the feature of matching the MSR by construction. Although it may be possible to find such a correlation that reduces $s_{FP}$, this method lacks both the theoretical motivation of the DW15 approach and the simplicity of the empirical method presented here.  This is worth exploring but we leave it for further work.

We note that previous semi-empirical models of elliptical galaxy dynamics such as those of~\citet{Dutton_1},~\citet{TG} and~\citet{Dutton_13} do not incorporate significant correlations between galaxy and halo properties beyond those investigated here. If we are right that matching the FP scatter requires such a correlation, it follows that these models too would disagree with the FP in detail. Ultimately, hydrodynamical simulations may be necessary to identify and understand all dynamically-relevant correlations of galaxy and halo variables.

\section{Conclusion}
\label{sec:conclusion}

The Fundamental Plane (FP) and Faber--Jackson relation (FJR) of elliptical galaxies provide an important theoretical testing ground, and contain information on galaxy formation processes unlikely to be found elsewhere. We have isolated five key characteristics of these relations and investigated the extent to which they are expected on the basis of a generalised abundance matching model in concordance $\Lambda$CDM. Comparing to a large and homogeneously-selected sample of elliptical galaxies from the NASA-Sloan Atlas, our main findings are the following.

\begin{enumerate}

\item{} The magnitude of observed velocity dispersions indicates on average a small amount of dark matter within the central regions of ellipticals with $M_* \approx 10^{10} \: \mathrm{M_\odot}$. The discrepancy with the prediction of a baryon-only model rises at higher stellar mass, reaching $\sim 0.15$ dex at $M_* = 10^{11.5} \: \mathrm{M_\odot}$.

\item{} Situating galaxies within the haloes produced by an $N$-body simulation using currently-favoured cosmological parameter values causes an overprediction of their velocity dispersions, unless dark matter is removed from the central regions. In the context of our framework, successfully matching the normalisation of the FJR requires a degree of halo expansion in response to disc formation similar to that required by~\citet{DW} for agreement with the Tully--Fisher relation. 

\item{} Abundance matching naturally predicts an FJR slope shallower than that observed. This discrepancy may potentially be accounted for by a variation in either effective radial orbit anisotropy or halo expansion factor with stellar mass, or may indicate non-universality in the IMF.

\item{} Across the parameter space spanned by our framework, the scatter predicted around the FP is significantly too high compared to the observations. This may indicate the presence of correlations between galaxy and halo variables not included in our models.

\item{} The rate at which velocity dispersion varies with size at fixed $M_*$ is a measure of the tilt of the FP. We demonstrate that the observed non-homology of elliptical galaxies (in this case, the variation of S\'{e}rsic index with $M_*$ and $R_\mathrm{eff}$), coupled with the variations in dynamical mass-to-light ratio generated by abundance matching, are sufficient to account for this key statistic. This is true in detail, however, only when the anisotropy of stellar motions is accounted for.

\end{enumerate}

There are several directions in which this work could be taken. First, it is important to understand the correlation between galaxy size and halo properties implied by the small scatter in the Fundamental Plane. This could be achieved either by exploring the output of hydrodynamical simulations which produce galaxies of realistic size, or by creating toy models in which correlations can be varied and constrained manually.  Second, the parameters of our framework have implications for various statistics of the large-scale galaxy distribution, including galaxy clustering and galaxy-galaxy lensing, and combining such analyses with dynamical modelling in a single self-consistent framework may be expected to considerably reduce the remaining uncertainties. Finally, in this work and~\citet{DW} we have treated early and late-type galaxies separately; future tests of galaxy formation models should do both together.

\section*{Acknowledgments}

We thank Aaron Dutton, Simon Foreman and Yao-Yuan Mao for comments on the manuscript. We further thank the anonymous referee for comments which improved the paper.  This work received partial support from the U.S.\ Department of Energy under contract number DE-AC02-76SF00515.  This research made use of the Dark Sky Simulations, which were produced using an INCITE 2014 allocation on the Oak Ridge Leadership Computing Facility at Oak Ridge National Laboratory.  We thank Sam Skillman, Mike Warren, Matt Turk and our other Dark Sky collaborators for their efforts in creating these simulations and for providing access to them.  Additional computation was performed at SLAC National Accelerator Laboratory.  We thank the SLAC computational team for their consistent support.

This paper made use of data from the NASA-Sloan Atlas (\url{www.nsatlas.org}).  We thank Michael Blanton for making this resource available and for his help in using it. This paper also made use of data from the Sloan Digital Sky Survey.  Funding for SDSS-III has been provided by the Alfred P. Sloan Foundation, the Participating Institutions, the National Science Foundation, and the U.S. Department of Energy. The SDSS-III web site is \url{www.sdss3.org}.  SDSS-III is managed by the Astrophysical Research Consortium for the Participating Institutions of the SDSS-III Collaboration including the University of Arizona, the Brazilian Participation Group, Brookhaven National Laboratory, University of Cambridge, University of Florida, the French Participation Group, the German Participation Group, the Instituto de Astrofisica de Canarias, the Michigan State/Notre Dame/JINA Participation Group, Johns Hopkins University, Lawrence Berkeley National Laboratory, Max Planck Institute for Astrophysics, New Mexico State University, New York University, Ohio State University, Pennsylvania State University, University of Portsmouth, Princeton University, the Spanish Participation Group, University of Tokyo, University of Utah, Vanderbilt University, University of Virginia, University of Washington, and Yale University.

\bsp	


\begin{thebibliography}{10}

\bibitem[Bamford et al.(2009)]{Bamford}
Bamford S.~P. et al., 2009, MNRAS, 393, 1324 

\bibitem[\protect\citeauthoryear{Behroozi, Conroy \& Wechsler}{Behroozi et al.}{2010}]{Behroozi_2010}
Behroozi P.~S., Conroy C., Wechsler R.~H., 2010, ApJ, 717, 379 

\bibitem[\protect\citeauthoryear{Behroozi, Wechsler \& Wu}{Behroozi et al.}{2013}]{Rockstar}
Behroozi P.~S., Wechsler R.~H., Wu H.-Y., 2013, ApJ, 762, 109 

\bibitem[\protect\citeauthoryear{Bender, Burstein \& Faber}{Bender et al.}{1992}]{Bender}
Bender R., Burstein D., Faber S.~M., 1992, ApJ, 399, 462 

\bibitem[Bernardi et al.(2003a)]{Bernardi_sample}
Bernardi M. et al., 2003, AJ, 125, 1817 

\bibitem[Bernardi et al.(2003b)]{Bernardi_FP}
Bernardi M. et al., 2003, AJ, 125, 1866 

\bibitem[Bernardi et al.(2013)]{Bernardi_SMF}
Bernardi M., Meert A., Sheth R.~K., Vikram V., Huertas-Company M., Mei F., Shankar F., 2013, MNRAS, 436, 697

\bibitem[Blanton \& Roweis(2007)]{Blanton_Roweis}
Blanton M.~R., Roweis S., 2007, AJ, 133, 734 

\bibitem[Blanton et al.(2011)]{Blanton_NSA}
Blanton M.~R., Kazin E., Muna D., Weaver B.~A., Price-Whelan A., 2011, AJ, 142, 31 

\bibitem[\protect\citeauthoryear{Blumenthal et al.}{1986}]{Blumenthal}
Blumenthal G.R., Faber S.M., Flores R., Primack J.R., 1986, ApJ, 301, 27

\bibitem[\protect\citeauthoryear{Borriello, Salucci \& Danese}{Borriello et al.}{2003}]{Borriello}
Borriello A., Salucci P., Danese L., 2003, MNRAS, 341, 1109 

\bibitem[\protect\citeauthoryear{Boylan-Kolchin, Bullock \& Kaplinghat}{Boylan-Kolchin et al.}{2011}]{TBTF}
Boylan-Kolchin M., Bullock J.~S., Kaplinghat M., 2011, MNRAS, 415, L40 

\bibitem[Chae \& Gong(2015)]{MOND_FP_baby}
Chae K.~H., Gong I.~T., 2015, MNRAS, 451, 1719

\bibitem[Cappellari(2015)]{Cappellari}
Cappellari M., 2015, IAU Symposium, 311, 20 

\bibitem[Cappellari(2016)]{Cappellari_review}
Cappellari M., 2016, Ann. Rev. Astron. Astrophys., 54, 597

\bibitem[Cappellari et al.(2007)]{Cappellari_anisotropy}
Cappellari M. et al., 2007, MNRAS, 379, 418 

\bibitem[Cappellari et al.(2013)]{Cappellari_2}
Cappellari M. et al., 2013, MNRAS, 432, 1709 

\bibitem[Cardone et al.(2011)]{MOND_FP}
Cardone V.~F., Angus G., Diaferio A., Tortora C., Molinaro R., 2011, MNRAS, 412, 2617 

\bibitem[Ciotti \& Bertin(1999)]{Ciotti}
Ciotti L., Bertin G., 1999, A\&A, 352, 447 

\bibitem[\protect\citeauthoryear{Conroy, Wechsler \& Kravtsov}{Conroy et al.}{2006}]{Conroy}
Conroy C., Wechsler R.~H., Kravtsov A.~V., 2006, ApJ, 647, 201 

\bibitem[Di Cintio et al.(2014)]{diCintio}
Di Cintio A., Brook C.~B., Dutton A.~A., Macci{\`o} A.~V., Stinson G.~S., Knebe A., 2014, MNRAS, 441, 2986 

\bibitem[Diemer \& Kravtsov(2015)]{Diemer_Kravtsov}
Diemer B., Kravtsov A.~V., 2015, ApJ, 799, 108

\bibitem[de Blok(2010)]{cusp-core}
de Blok W.~J.~G., 2010, Adv. Astron., 2010, 789293

\bibitem[Desmond \& Wechsler(2015)]{DW}
Desmond H., Wechsler R.~H., 2015, MNRAS, 454, 322 

\bibitem[Djorgovski \& Davis(1987)]{Djorgovski}
Djorgovski S., Davis M., 1987, ApJ, 313, 59 

\bibitem[Dressler et al.(1987)]{Dressler}
Dressler A., Lynden-Bell D., Burstein D., Davies R., Faber S., Terlevich R., Wegner G., 1987, ApJ, 313, 42 

\bibitem[\protect\citeauthoryear{D'Souza, Vegetti \& Kauffman}{D'Souza et al.}{2015}]{SMF_latest}
D'Souza R., Vegetti S., Kauffmann G., 2015, MNRAS, 454, 4027 

\bibitem[Dutton et al.(2007)]{D07}
Dutton A.~A., van den Bosch F.~C., Dekel A., Courteau S., 2007, ApJ, 654, 27 

\bibitem[Dutton et al.(2011)]{Dutton_1}
Dutton A.~A. et al., 2011, MNRAS, 416, 322  

\bibitem[Dutton et al.(2013)]{Dutton_13}
Dutton A.~A., Macci{\`o} A.~V., Mendel J.~T., Simard L., 2013, MNRAS, 432, 2496 

\bibitem[Faber \& Jackson(1976)]{FJ}
Faber S.~M., Jackson R.~E., 1976, ApJ, 204, 668 

\bibitem[Gentile et al.(2009)]{Gentile}
Gentile G., Famaey B., Zhao H., Salucci P., 2009, Nature, 461, 627 

\bibitem[Gerhard et al.(2001)]{Gerhard_anisotropy}
Gerhard O., Kronawitter A., Saglia R.~P., Bender R., 2001, AJ, 121, 1936 

\bibitem[Gnedin et al.(2004)]{Gnedin_2004}
Gnedin O.~Y., Kravtsov A.~V., Klypin A.~A., Nagai D., 2004, ApJ, 616, 16

\bibitem[\protect\citeauthoryear{Gnedin et al.}{2011}]{Gnedin_2011}
Gnedin O.Y., Ceverino D., Gnedin N.Y., Klypin A.A., Kravtsov A.V., Levine R., Nagai D., Yepes G., 2011, preprint (arXiv:1108.5736)

\bibitem[Guo et al.(2010)]{Guo}
Guo Q., White S., Li C., Boylan-Kolchin M., 2010, MNRAS, 404, 1111

\bibitem[Hearin et al.(2014)]{CAM}
Hearin A.~P., Watson D.~F., Becker M.~R., Reyes R., Berlind A.~A., Zentner A.~A., 2014, MNRAS, 444, 729 

\bibitem[Henriques et al.(2015)]{Henriques}
Henriques B.~M.~B., White S.~D.~M., Thomas P.~A., Angulo R., Guo Q., Lemson G., Springel V., Overzier R, 2015, MNRAS, 451, 2663 

\bibitem[Khandai et al.(2015)]{MassiveBlack}
Khandai N., Di Matteo T., Croft R., Wilkins S., Feng Y., Tucker E., DeGraf C., Liu M., 2015, MNRAS 450, 1349 

\bibitem[Kravtsov et al.(2004)]{Kravtsov}
Kravtsov A.~V., Berlind A.~A., Wechsler R.~H., Klypin A.~A., Gottlober S., Allgood B., Primack J.~R., 2004, ApJ, 609, 35

\bibitem[Kravtsov(2013)]{Kravtsov13}
Kravtsov A.~V., 2013, ApJ, 764, L31

\bibitem[Lehmann et al.(2015)]{Lehmann}
Lehmann B.~V., Mao Y.-Y., Becker M.~R., Skillman S.~W., Wechsler R.~H., 2015, preprint (arXiv:1510.05651)

\bibitem[\protect\citeauthoryear{Lelli, McGaugh \& Schombert}{Lelli et al.}{2016}]{Lelli}
Lelli F., McGaugh S.~S., Schombert, J~M., 2016, ApJ, 816, L14 

\bibitem[Lintott et al.(2008)]{Lintott}
Lintott C.~J. et al., 2008, MNRAS, 389, 1179

\bibitem[Mamon \& {\L}okas(2005)]{Mamon_Lokas}
Mamon G.~A., {\L}okas E.~L., 2005, MNRAS, 363, 705

\bibitem[Mandelbaum et al.(2016)]{Mandelbaum}
Mandelbaum R., Wang W., Zu Y., White S., Henriques B., More S., 2016, MNRAS, 457, 3200 

\bibitem[Moster et al.(2010)]{Moster}
Moster B.~P., Somerville R.~S., Maulbetsch C., van den Bosch F.~C., Macci{\`o} A.~V., Naab T., Oser L., 2010, ApJ, 710, 903 

\bibitem[Macci{\`o} et al.(2012)]{HydroSim_cores}
Macci{\`o} A.~V., Stinson G., Brook C.~B., Wadsley J., Couchman H.~M.~P., Shen S., Gibson B.~K., Quinn T., 2012, ApJ, 744, L9 

\bibitem[Mar{\'{\i}}n et al.(2008)]{Marin08}
Mar{\'{\i}}n F.~A., Wechsler R.~H., Frieman J.~A., Nichol R.~C., 2008, ApJ,  672, 849 

\bibitem[McGaugh(2004)]{McGaugh_MDA}
McGaugh S.~S., 2004, ApJ, 609, 652 

\bibitem[\protect\citeauthoryear{Mo, Mao \& White}{Mo et al.}{1998}]{MMW}
Mo H.J., Mao S., White S.D.M., 1998, MNRAS, 295, 319

\bibitem[Napolitano et al.(2009)]{Napolitano}
Napolitano et al., 2009, MNRAS, 393, 329 

\bibitem[Nigoche-Netro et al.(2015)]{Nigoche}
Nigoche-Netro A., Ruelas-Mayorga A., Lagos P., Ramos-Larios G., Kehrig C., Kemp S.~N., Montero-Dorta A.~D., Gonzalez-Cervantes J., 2015, MNRAS, 446, 85

\bibitem[Pizagno et al.(2007)]{P07}
Pizagno J. et al., 2007, AJ, 134, 945 (P07)

\bibitem[Pontzen \& Governato(2014)]{Pontzen}
Pontzen A., Governato F., 2014, Nature, 506, 171 

\bibitem[Prugniel \& Simien(1997)]{Prugniel_Simien}
Prugniel P., Simien F., 1997, A\&A, 321, 111 

\bibitem[Reddick et al.(2013)]{Reddick}
Reddick R.~M., Wechsler R.~H., Tinker J.~L., Behroozi P.~S., 2013, ApJ, 771, 30 

\bibitem[Romanowsky et al.(2003)]{Romanowsky}
Romanowsky A.~J., Douglas N.~G., Arnaboldi M., Kuijken K., Merrifield M.R., Napolitano N.R., Capaccioli M., Freeman K.C, 2003, Science, 301, 1696 

\bibitem[Rodr{\'{\i}}guez-Puebla et al.(2011)]{Puebla} 
Rodr{\'{\i}}guez-Puebla A., Avila-Reese V., Firmani C., Col{\'{\i}}n P., 2011, Rev. Mex. Astron. Astrofis., 47, 235 

\bibitem[Salucci(2004)]{Salucci}
Salucci P., 2004, Particle Physics Beyond the Standard Model, 92, 613 

\bibitem[Sanders(1990)]{Sanders_MDA}
Sanders R.~H., 1990, A\&AR, 2, 1 

\bibitem[Sanders(2000)]{Sanders_FP}
Sanders R.~H., 2000, MNRAS, 313, 767

\bibitem[Sanders(2010)]{Sanders_FJR}
Sanders R.~H., 2010, MNRAS, 407, 1128

\bibitem[Schaye et al.(2015)]{EAGLE}
Schaye J. et al., 2015, MNRAS, 446, 521

\bibitem[\protect\citeauthoryear{Schulz, Mandelbaum, \& Padmanabhan}{Schulz et al.}{2010}]{Schulz}
Schulz A.~E., Mandelbaum R., Padmanabhan N., 2010, MNRAS, 408, 1463 

\bibitem[S\'{e}rsic(1968)]{Sersic_1968}
S\'{e}rsic J.~L., 1968, Cordoba, Argentina: Observatorio Astronomico

\bibitem[Skillman et al.(2014)]{DarkSky}
Skillman S.~W., Warren M.~S., Turk M.~J., Wechsler R.~H., Holz D.~E., Sutter P.~M., 2014, preprint (arXiv:1407.2600)

\bibitem[Sonnenfeld et al.(2015)]{Sonnenfeld}
Sonnenfeld A., Treu T., Marshall P.~J., Suyu S.~H., Gavazzi R., Auger M.~W., Nipoti C., 2015, ApJ, 800, 94

\bibitem[Thomas et al.(2007)]{Thomas_anisotropy}
Thomas J., Saglia R.~P., Bender R., Thomas D., Gebhardt K., Magorrian J., Corsini E.~M., Wegner G., 2007, MNRAS, 382, 657 

\bibitem[Tiret \& Combes(2009)]{Tiret_Combes}
Tiret O., Combes F., 2009, A\&A, 496, 659

\bibitem[Tiret et al.(2007)]{Tiret_Combes_MOND_VD}
Tiret O., Combes F., Angus G.~W., Famaey B., Zhao H.~S., 2007, A\&A, 476, L1 

\bibitem[Trujillo-Gomez et al.(2011)]{TG}
Trujillo-Gomez S., Klypin A., Primack J., Romanowsky A.~J., 2011, ApJ, 742, 16 

\bibitem[van Uitert et al.(2016)]{Uitert}
van Uitert E. et al., 2016, MNRAS,  459, 3251

\bibitem[Vogelsberger et al.(2014)]{Illustris}
Vogelsberger M. et al., 2014, MNRAS, 444, 1518 

\bibitem[Warren(2013)]{Warren13}
Warren M.~S., 2013, Proc. Int. Conf. High Perform. Comput. Netw. Storage Anal., (New York: ACM), p. 72

\bibitem[Wojtak \& Mamon(2013)]{Wojtak}
Wojtak R., Mamon G.~A., 2013, MNRAS, 428, 2407 

\bibitem[Zehavi et al.(2011)]{Zehavi}
Zehavi I. et al., 2011, ApJ, 736, 59 

\end{thebibliography}
\end{document}